\documentclass[journal]{IEEEtran}

\usepackage[noadjust]{cite}

\usepackage{graphicx}

\usepackage[cmex10]{amsmath}
\usepackage{amssymb}
\usepackage{bm}
\usepackage{cases}

\newtheorem{theorem}{Theorem}[section]
\newtheorem{lemma}[theorem]{Lemma}

\newtheorem{definition}[theorem]{Definition}

\newtheorem{remark}[theorem]{Remark}

\newcommand{\relvar}[2]{\buildrel {#2} \over {#1}}
\newcommand{\eqvar}[1]{\relvar{=}{\mathrm{#1}}}
\newcommand{\ltvar}[1]{\relvar{<}{\mathrm{#1}}}

\newcommand{\levar}[1]{\relvar{\le}{\mathrm{#1}}}

\newcommand{\eqdef}{\eqvar{\scriptscriptstyle\triangle}}

\newcommand{\findent}{\hspace{1em}}
\newlength{\nullrellength}
\newcommand{\nullrel}{\:\hspace{\nullrellength}\:}
\newcommand{\breakop}[1]{\hspace{-0.2222em}{}#1}

\newcommand{\integers}{\mathbb{Z}}
\newcommand{\pintegers}{\mathbb{N}}
\newcommand{\nnintegers}{\mathbb{N}_0}
\newcommand{\realnumbers}{\mathbb{R}}

\newcommand{\unitsof}[1]{{#1^{\times}}}

\newcommand{\field}[1]{\mathbb{F}_{#1}}
\newcommand{\fieldstar}[1]{\unitsof{\field{#1}}}

\newcommand{\symmetricgroup}[1]{\mathrm{S}_{#1}}

\newcommand{\seq}[1]{\mathbf{#1}}

\DeclareMathOperator{\order}{\Theta}

\newcommand{\floor}[1]{\left\lfloor {#1} \right\rfloor}
\newcommand{\ceil}[1]{\left\lceil {#1} \right\rceil}

\newcommand{\pprod}{\odot}
\newcommand{\bigpprod}{\bigodot}
\DeclareMathOperator{\coef}{coef}
\DeclareMathOperator{\dmin}{d_{min}}

\newcommand{\codevar}[2]{f^{\mathrm{#1}}_{#2}}
\newcommand{\rcodevar}[2]{F^{\mathrm{#1}}_{#2}}
\newcommand{\rmcode}[1]{\rcodevar{RM}{#1}}
\newcommand{\ldcode}[1]{\rcodevar{LD}{#1}}
\newcommand{\repcode}[1]{\codevar{REP}{#1}}

\newcommand{\chkcode}[1]{\codevar{CHK}{#1}}

\begin{document}

\title{Weight Distributions of Regular Low-Density Parity-Check Codes over Finite Fields}
\author{Shengtian~Yang,~\IEEEmembership{Member,~IEEE,} Thomas~Honold,~\IEEEmembership{Member,~IEEE,} Yan~Chen,~\IEEEmembership{Member,~IEEE,} Zhaoyang~Zhang,~\IEEEmembership{Member,~IEEE,} Peiliang~Qiu,~\IEEEmembership{Member,~IEEE}
\thanks{This work was supported in part by the National Natural Science Foundation of China under Grants 60772093, 60802014, and 60872063, the Chinese Specialized Research Fund for the Doctoral Program of Higher Education under Grants 200803351023 and 200803351027, the National High Technology Research and Development Program of China under Grant 2007AA01Z257, the Zhejiang Provincial Natural Science Foundation of China under Grant Y106068, and the Program for New Century Excellent Talents in University under grant NCET-09-0701.}
\thanks{S. Yang is self-employed at Zhengyuan Xiaoqu 10-2-101, Hangzhou 310011, China (email: yangst@codlab.net).}
\thanks{T. Honold is with the Department of Information Science and Electronic Engineering, Zhejiang University, Hangzhou 310027, China (email: \mbox{honold@zju.edu.cn}).}
\thanks{Y. Chen is with Huawei Technologies Co., Ltd (Shanghai), Shanghai 201206, China (email: eeyanchen@huawei.com).}
\thanks{Z. Zhang and P. Qiu are with the Department of Information Science and Electronic Engineering, Zhejiang University, Hangzhou 310027, China (email: ning\_ming@zju.edu.cn; qiupl@zju.edu.cn).}
\thanks{Copyright (c) 2011 IEEE. Personal use of this material is permitted.}
}

\markboth{Accepted for Publication in IEEE Transactions on Information Theory (Version: \today)}{}


\maketitle

\begin{abstract}
The average weight distribution of a regular low-density parity-check (LDPC) code ensemble over a finite field is thoroughly analyzed. In particular, a precise asymptotic approximation of the average weight distribution is derived for the small-weight case, and a series of fundamental qualitative properties of the asymptotic growth rate of the average weight distribution are proved. Based on this analysis, a general result, including all previous results as special cases, is established for the minimum distance of individual codes in a regular LDPC code ensemble.
\end{abstract}


\IEEEpeerreviewmaketitle

\section{Introduction}\label{sec:Introduction}

\IEEEPARstart{L}{ow-density} parity-check (LDPC) codes, originally introduced by Gallager \cite{JSCC:Gallager196300}, are a family of linear codes characterized by a sparse parity-check matrix. Owing to their capacity-approaching performance under low-complexity iterative decoding algorithms, LDPC codes have attracted tremendous attention in the past years. To evaluate the theoretical performance of an LDPC code, a typical method is to estimate its performance under maximum-likelihood (ML) or iterative decoding assumptions. The performance of a linear code under ML decoding can be well estimated based on its weight distribution \cite{JSCC:Gallager196300}, so having the knowledge about weight distributions of LDPC codes facilitate the analysis of the ML decoding performance.

The first analysis work on the weight distributions of LDPC codes was given by Gallager in his pioneering work \cite{JSCC:Gallager196300}, where he studied the weight distributions of binary regular LDPC codes. Moreover, he also generalized the analysis to non-binary regular LDPC codes over $\integers_m$ ($m>2$), characterized by zero-one parity-check matrices. Ever since the publication of \cite{JSCC:Gallager196300}, there has been a lot of work extending the analysis of weight distributions of binary LDPC codes in different ways, such as \cite{JSCC:Litsyn200204, JSCC:Litsyn200312, JSCC:Burshtein200406, JSCC:Di200611, JSCC:Rathi200609, JSCC:Flanagan200911}. A generalization of weight distributions, also known as spectra, of regular LDPC codes over finite fields and arbitrary abelian groups were later studied in \cite{JSCC:Bennatan200403, JSCC:Como2008}. More recently, the binary weight distributions of non-binary LDPC codes also received some attention \cite{JSCC:Andriyanova200906}. By now a bundle of formulas about weight distributions of various LDPC codes is known, but the value and significance of most formulas is far from being fully understood, except in the case of binary regular LDPC codes, which have been well studied \cite{JSCC:Gallager196300, JSCC:Litsyn200204}. The difficulty is due to the complex expressions for the weight distributions of LDPC codes, which are usually obtained by the generating function approach and hence are typically expressed as coefficients of a polynomial. Given a polynomial $p(x)$ with nonnegative coefficients, a usual approach for estimating the coefficient of a monomial $x^k$ in $[p(x)]^n$ is to calculate the infimum of $[p(x)]^n/x^k$ over all positive $x$, which gives an upper bound of the coefficient and in fact has the same asymptotic growth rate as the coefficient \cite[Theorem~1]{JSCC:Burshtein200406}. However, analyzing functions like $\inf_{y>0} f(x,y)$ is not an easy job. When $f(x,y)$ is complicated, determining the shape, such as monotonicity, convexity, and zeros, of $\inf_{y>0} f(x,y)$ becomes a difficult mission.

In this paper, we shall perform such a mission for ensembles of regular LDPC codes over finite fields. At first, as an easy consequence of the results in \cite{JSCC:Bennatan200403, JSCC:Como2008, JSCC:Yang200909}, an exact expression is introduced for the average weight distribution of a $(c,d)$-regular LDPC code ensemble over the finite field $\field{q}$ of order $q$, where $c$ and $d$, in a less strict sense, correspond to the column and row weight of parity-check matrix, respectively. Based on this expression, we show that, when averaged on the whole ensemble, the fraction of codewords of small weight $l$ in an LDPC code is at most asymptotically $n^{-\ceil{(c-2)l/2}}$ as the coding length $n$ goes to infinity. Next, using the upper-bound technique mentioned above, we analyze the asymptotic growth rate $\omega_{q,c,d}(x)$ of the average weight distribution, where $x$ denotes the normalized weight. A series of fundamental qualitative properties of $\omega_{q,c,d}(x)$ are found and proved. In particular, we show that for $d \ge c \ge 3$, $\omega_{q,c,d}(x)$ has a unique zero $x_0$ in $(0, 1-1/q]$. This zero just corresponds to the normalized minimum distance of a typical LDPC code, and hence provides important information about the code ensemble. Finally, we prove that for $d\ge c\ge 3$, there are at most a fraction $\order(n^{-\ceil{(c-2)l_0/2}})$ of all codes in the ensemble whose minimum distance is between the constant $l_0$ and $\alpha n$, where $\alpha \in (0, x_0)$.

The rest of this paper is organized as follows. In Section~\ref{sec:NotationsAndConventions}, we introduce the notations and conventions to be used throughout the paper. In Section~\ref{sec:WeightDistributionOfLDPCCodes}, we define the ensemble of regular LDPC codes over a finite field and give its average weight distribution function; moreover, we study the asymptotic behavior of the average weight distribution for the small-weight case. The main analysis, consisting of two stages, for the asymptotic growth rate of the average weight distribution is performed in Sections~\ref{sec:Function1} and \ref{sec:Function2}. The minimum distance of individual codes in a regular LDPC code ensemble is analyzed in Section~\ref{sec:MinimumDistance}. Section~\ref{sec:Conclusion} concludes the paper.

\section{Notations and Conventions}\label{sec:NotationsAndConventions}

In this section, we introduce some basic notations and conventions to be used throughout the rest of this paper.

\begin{itemize}
\item In general, symbols, real variables, and deterministic mappings are denoted by lowercase letters. Sets and random elements are denoted by capital letters.

\item The symbols $\integers$, $\pintegers$, $\nnintegers$, $\realnumbers$ denote the ring of integers, the set of positive integers, the set of nonnegative integers, and the field of real numbers, respectively. For a prime power $q\ge 2$ the finite field of order $q$ is denoted by $\field{q}$. The multiplicative subgroup of nonzero elements of $\field{q}$ is denoted by $\fieldstar{q}$.

\item The $n$-fold cartesian product of a set $A$ is denoted by $A^n$. An element of $A^n$ is denoted by $\seq{x} = (x_1, x_2, \ldots, x_n)$, where $x_i \in A$ denotes the $i$th component of $\seq{x}$.

\item For any vector $\seq{c} \in \field{q}^n$, the \emph{weight} $w(\seq{c})$ of $\seq{c}$ is the number of nonzero symbols in it, that is, $w(\seq{c})\eqdef |\{i : c_i\ne 0\}|$.

\item Given the functions $f: X \to Y$ and $g: Y \to Z$, their composite is the function $g \circ f: X \to Z$ given by $x \mapsto g(f(x))$.

\item Given the functions $f: X_1 \to Y_1$ and $g: X_2 \to Y_2$, their cartesian product is the function $f \pprod g: X_1 \times X_2 \to Y_1 \times Y_2$ given by $(x_1, x_2) \mapsto (f(x_1), g(x_2))$.

\item When performing probabilistic analysis, all objects of study are relative to a basic probability space $(\Omega, \mathcal{A}, P)$ where $\mathcal{A}$ is a $\sigma$-algebra in $\Omega$ and $P$ is a probability measure on $(\Omega, \mathcal{A})$. For any event $A \in \mathcal{A}$, $PA = P(A)$ is called the probability of $A$. Any measurable mapping of $\Omega$ into some measurable space $(B, \mathcal{B})$ is generally called a random element. For any random set or function, we tacitly assume that their $n$-fold cartesian products (e.g., $A^n$ or $\bigpprod_{i=1}^n F$) are cartesian products of their independent copies.
\item All logarithms are taken to the natural base $\mathrm{e}$ and denoted by $\ln$.

\item For any $x \in [0,1]$ and any integer $q \ge 2$, the \emph{entropy function} $H_q(x)$ is defined by
\[
H_q(x) \eqdef x\ln\frac{1}{x} + (1-x)\ln\frac{1}{1-x} + x \ln(q-1).
\]
For any $x, y \in [0, 1]$, the \emph{information divergence function} $D(x\|y)$ is defined by
$$
D(x\|y) \eqdef x\ln\frac{x}{y} + (1-x)\ln\frac{1-x}{1-y}.
$$

\item For any real functions $f(n)$ and $g(n)$ with $n \in \pintegers$, the asymptotic $\Theta$-notation $f(n)=\order(g(n))$ means that there exist positive constants $c_1$ and $c_2$ such that
\[
c_1g(n) \le f(n) \le c_2g(n).
\]
for sufficiently large $n$.

\item For $x\in\realnumbers$, $\floor{x}$ denotes the largest integer not exceeding $x$, and $\ceil{x}$ denotes the smallest integer not less than $x$.
\end{itemize}

\section{Regular LDPC Codes over Finite Fields}\label{sec:WeightDistributionOfLDPCCodes}

We first define some basic $\field{q}$-linear transformations.

\begin{definition}
A \emph{single symbol repetition} with a parameter $c \in \pintegers$ is a mapping $\repcode{q,c}: \field{q} \to \field{q}^c$ given by $x \mapsto (x, x, \ldots, x)$.
\end{definition}

\begin{definition}
A \emph{single symbol check} with a parameter $d \in \pintegers$ is a mapping  $\chkcode{q,d}: \field{q}^d \to \field{q}$ given by $\seq{x} \mapsto \sum_{i=1}^d x_i$.
\end{definition}

\begin{definition}\label{df:RandomMultiplier}
A \emph{single symbol random multiplier map} is a random mapping $\rmcode{q}: \field{q} \to \field{q}$ given by $x \to Cx$ where $C$ is an independent random variable uniformly distributed over $\fieldstar{q}$.
\end{definition}

\begin{definition}\label{df:RandomInterleaver}
A \emph{uniform random interleaver} of $\field{q}^n$ is a random automorphism $\Sigma_{q,n}: \field{q}^n \to \field{q}^n$ given by $\seq{x} \mapsto (x_{\Pi^{-1}(1)}, x_{\Pi^{-1}(2)}, \ldots, x_{\Pi^{-1}(n)})$, where $\Pi$ is an independent random permutation uniformly distributed over the symmetric group $\symmetricgroup{n}$, i.e., all permutations on $n$ letters.
\end{definition}

Next, we define a random linear transformation based on the above simple maps.

\begin{definition}
$\ldcode{q,c,d,n}: \mathbb{F}_q^{n} \to \mathbb{F}_q^{cn/d}$ is a random mapping defined by
\begin{equation}\label{eq:DefinitionOfLDGMCode}
\ldcode{q,c,d,n} \eqdef \chkcode{q,d,cn/d} \circ \rmcode{q,cn} \circ \Sigma_{q,cn} \circ \repcode{q,c,n}
\end{equation}
where $c,d \in \pintegers$, $d$ divides $cn$, and
\[
\repcode{q,c,n} \eqdef \bigpprod_{i=1}^n \repcode{q,c},\;
\chkcode{q,d,n} \eqdef \bigpprod_{i=1}^{n} \chkcode{q,d},\;
\rmcode{q,n} \eqdef \bigpprod_{i=1}^n \rmcode{q}.
\]
\end{definition}

Considering the kernel of $\ldcode{q,c,d,n}$, we thus obtain an ensemble of regular LDPC codes over $\field{q}$, which is called a \emph{random $(c,d)$-regular LDPC code over $\field{q}$} and is denoted by $\mathcal{C}_{q,c,d}^{(n)}$.\footnote{We shall tacitly assume throughout the paper that the block length $n$ always takes values such that $d$ divides $cn$.} This ensemble was originally introduced in \cite{JSCC:Luby200102,JSCC:Bennatan200403,JSCC:Richardson200102} by the method of bipartite graphs.

To see the connection of $\ldcode{q,c,d,n}$ with a bipartite graph, we
may regard each $\repcode{q,c}$ as a variable node with $c$ sockets
and each $\chkcode{q,d}$ as a check node with $d$ sockets. Then in
total there are $nc$ variable sockets and $nc$ check sockets. We say
that the $i$th variable socket  and the $j$th check socket are
connected by an edge if $j = \Pi(i)$, where $\Pi$ is the
random permutation defined in
Definition~\ref{df:RandomInterleaver}. We also define the label
  of the edge connecting these two sockets to be the random variable
  $C$ defined in Definition~\ref{df:RandomMultiplier}. 
  Then we dispose of the sockets (i.e.\ edges are considered as
  connections between variable nodes and check nodes). The resulting
  random graph (which may have repeated edges) is exactly the
  random regular bipartite graph with
  independent and uniformly distributed random edge labels taken from
  $\fieldstar{q}$ as in \cite{JSCC:Bennatan200403}.

Now let us investigate the weight distribution of $\mathcal{C}_{q,c,d}^{(n)}$. The next theorem gives its average weight distribution.

\begin{theorem}[cf. \cite{JSCC:Bennatan200403, JSCC:Como2008, JSCC:Yang200909}]\label{th:LDPCWeightDistribution}
For $c,d\in \pintegers$, the average weight distribution of $\mathcal{C}_{q,c,d}^{(n)}$ is given by
\begin{equation}
E{\left[A_{q,c,d}^{(n)}(l)\right]} = \frac{{n \choose l} \coef{\left(g_{q,d}^{(cn/d)}(x), x^{cl}\right)}}{{cn \choose cl} (q-1)^{(c-1)l}} \label{eq:LDPCWeightDistribution}
\end{equation}
where $A_{q,c,d}^{(n)}(l)$ denotes the number of codewords of weight $l$ in $\mathcal{C}_{q,c,d}^{(n)}$ ($0 \le l \le n$), $\coef{\left(p(x),x^l\right)}$ denotes the coefficient of $x^l$ in the polynomial $p(x)$, and
\begin{equation}
g_{q,d}^{(n)}(x) \eqdef \frac{1}{q^{n}} \left\{ [1+(q-1)x]^d + (q-1)(1-x)^d \right\}^{n}. \label{eq:DefinitionOfg_qd}
\end{equation}
Furthermore, we have
\begin{equation}
\frac{1}{n} \ln E{\left[A_{q,c,d}^{(n)}(l)\right]} \le \omega_{q,c,d}\left(\frac{l}{n}\right) + c\beta_{cn}(cl) \label{eq:LDPCWeightDistributionUpperBound}
\end{equation}
where
\begin{equation}
\omega_{q,c,d}(x) \eqdef H_q(x)+\frac{c}{d}\left[\delta_{q,d}(x)-\ln q\right] \label{eq:DefinitionOfOmega}
\end{equation}
\begin{equation}
\delta_{q,d}(x) \eqdef \inf_{\hat{x}\in(0,1)} \delta_{q,d}(x,\hat{x}) \label{eq:DefinitionOfDelta}
\end{equation}
\begin{equation}
\delta_{q,d}(x,\hat{x}) \eqdef dD(x\|\hat{x})+\rho_{q,d}(\hat{x}) \label{eq:DefinitionOfDelta2}
\end{equation}
\begin{equation}
\rho_{q,d}(x) \eqdef \ln\left[ 1 + (q - 1) \left( 1-\frac{qx}{q-1} \right)^d \right] \label{eq:DefinitionOfRho}
\end{equation}
\begin{equation}
\beta_{n}(l) \eqdef H_2{\left(\frac{l}{n}\right)}-\frac{1}{n}\ln{n \choose l}.
\end{equation}
\end{theorem}

\begin{IEEEproof}
The average weight distribution \eqref{eq:LDPCWeightDistribution} is in fact a known result. Note that
\[
E{\left[A_{q,c,d}^{(n)}(l)\right]}
= {n \choose l} (q-1)^l P{\left\{\left.\seq{c} \in \mathcal{C}_{q,c,d}^{(n)} \right| w(\seq{c}) = l\right\}}
\]
and
\[
P{\left\{\left.\seq{c} \in \mathcal{C}_{q,c,d}^{(n)} \right| w(\seq{c}) = l\right\}}
= \frac{\left|\left\{\hat{\seq{c}} \in \ker \chkcode{q,d,cn/d}: w(\hat{\seq{c}}) = cl\right\}\right|}{{cn \choose cl} (q-1)^{cl}}.
\]
For a proof of
\[
\left|\left\{\hat{\seq{c}} \in \ker \chkcode{q,d,cn/d}: w(\hat{\seq{c}}) = cl\right\}\right|
= \coef{\left(g_{q,d}^{(cn/d)}(x), x^{cl}\right)}
\]
the reader is referred to \cite[Appendix~III]{JSCC:Bennatan200403}, \cite{JSCC:Como2008, JSCC:Yang200909}. 

Now let us prove the inequality \eqref{eq:LDPCWeightDistributionUpperBound}. By the upper-bound technique introduced in Section~\ref{sec:Introduction}, it follows from \eqref{eq:LDPCWeightDistribution} that
\begin{IEEEeqnarray*}{rCl}
E{\left[A_{q,c,d}^{(n)}(l)\right]} \le \frac{{n \choose l} g_{q,d}^{(cn/d)}(x)}{{cn \choose cl} (q-1)^{(c-1)l} x^{cl}}
\end{IEEEeqnarray*}
for any $x > 0$. Taking
\[
x=\frac{\hat{x}}{(q-1)(1-\hat{x})}
\]
where $\hat{x} \in (0,1)$, we obtain
\begin{equation}
E{\left[A_{q,c,d}^{(n)}(l)\right]} \le \frac{(q-1)^l {n \choose l} \hat{g}_{q,d}^{(cn/d)}(\hat{x})}{{cn \choose cl} \hat{x}^{cl} (1-\hat{x})^{cn-cl}} \label{eq:Eq1InProofOfLDPCWDThm}
\end{equation}
where
\[
\hat{g}_{q,d}^{(n)}(x) \eqdef \frac{1}{q^n} \left[ 1 + (q-1)\left(1-\frac{qx}{q-1}\right)^d \right]^{n}.
\]
Taking logarithms of both sides of \eqref{eq:Eq1InProofOfLDPCWDThm} and using the lower-bound in Lemma~\ref{le:Inequality1}, we further have
\[
\frac{1}{n} \ln E{\left[A_{q,c,d}^{(n)}(l)\right]}
\le H_q(\alpha) + \frac{c}{d}\left[\delta_{q,d}(\alpha,\hat{x})-\ln q\right] + c\beta_{cn}(cl)
\]
where $\alpha\eqdef l/n$. The theorem is finally established by taking the infimum of the right side over all $\hat{x}\in(0,1)$.
\end{IEEEproof}

\begin{remark}
Loosely speaking, for any $\alpha \in [0,1]$, if we take $l=\alpha n$, then it follows from \cite[Theorem~1]{JSCC:Burshtein200406} that
\begin{IEEEeqnarray*}{rCl}
\lim_{n \to \infty} \frac{1}{n} \ln \coef{\left(g_{q,d}^{(cn/d)}(x), x^{cl}\right)}
&= &\frac{c}{d} \inf_{x>0} \ln \frac{g_{q,d}^{(1)}(x)}{x^{d\alpha}} \\
&= &\inf_{x>0} \frac{1}{m} \ln \frac{g_{q,d}^{(cm/d)}(x)}{x^{cm\alpha}}
\end{IEEEeqnarray*}
for any $m>0$. Comparing this identity with the proof of Theorem~\ref{th:LDPCWeightDistribution} and noting that the second term in the right hand side of \eqref{eq:LDPCWeightDistributionUpperBound} is asymptotically negligible, we immediately have
\[
\lim_{n \to \infty} \frac{1}{n} \ln E{\left[A_{q,c,d}^{(n)}(\alpha n)\right]} = \omega_{q,c,d}(\alpha).
\]
The function $\omega_{q,c,d}(x)$ thus represents the asymptotic growth rate of the average weight distribution of $\mathcal{C}_{q,c,d}^{(n)}$, and hence deserves further investigations. In the subsequent sections, we shall provide an in-depth analysis of $\omega_{q,c,d}(x)$.
\end{remark}

Although in general the average weight distribution of $\mathcal{C}_{q,c,d}^{(n)}$ is very complex, it becomes simple for some special $d$. The next two theorems give its complete characterization for $d=1,2$.

\begin{theorem}
\begin{numcases}{E{\left[A_{q,c,1}^{(n)}(l)\right]} =}
1 &$l=0$ \IEEEnonumber \\
0 &otherwise. \IEEEnonumber
\end{numcases}
\end{theorem}

\begin{IEEEproof}
For $d=1$, we have $\ldcode{q,c,d,n} = \rmcode{q,cn} \circ \Sigma_{q,cn} \circ \repcode{q,c,n}$, which is injective. In other words, the defining parity-check matrix of $\mathcal{C}_{q,c,1}^{(n)}$ has
rank $n$, so that $\mathcal{C}_{q,c,1}^{(n)}=\{\seq{0}\}$. 
\end{IEEEproof}

\begin{theorem}\label{th:LDPCWeightDistributionD2}
\begin{equation}
E{\left[A_{q,c,2}^{(n)}(l)\right]} =\left\{\begin{array}{ll}
\frac{{n \choose l} {cn/2 \choose cl/2}}{(q-1)^{(c/2-1)l} {cn \choose cl}} &\mbox{$cl$ is even} \\
0 &\mbox{otherwise}
\end{array}\right.\label{eq:LDPCWeightDistributionD2}
\end{equation}
\begin{equation}
\frac{1}{n} \ln E{\left[A_{q,c,2}^{(n)}(l)\right]} \le \left(1-\frac{c}{2}\right) H_q{\left(\frac{l}{n}\right)} + c\beta_{cn}(cl). \label{eq:LDPCWeightDistributionD2UpperBound}
\end{equation}
\end{theorem}

\begin{IEEEproof}
By \eqref{eq:DefinitionOfg_qd} it follows that
\[
g_{q,2}^{(n)}(x)  = \left[1+(q-1)x^2\right]^n.
\]
Then we have
\[
\coef{\left(g_{q,2}^{(cn/2)}(x), x^{cl}\right)} =
\left\{\begin{array}{ll}
\textstyle (q-1)^{cl/2} {cn/2 \choose cl/2} &\mbox{$cl$ is even} \\
0 &\mbox{otherwise.}
\end{array}\right.
\]
This together with \eqref{eq:LDPCWeightDistribution} gives \eqref{eq:LDPCWeightDistributionD2}, which further yields \eqref{eq:LDPCWeightDistributionD2UpperBound} by Lemma~\ref{le:Inequality1}.
\end{IEEEproof}

As shown above, the average weight distribution of $\mathcal{C}_{q,c,d}^{(n)}$ is trivial for $d=1,2$. In the sequel, we shall therefore concentrate on the general case of $d \ge 3$.

Another well-known fact to be noted is that when $q=2$ and $d$ is even, the weight distribution of $\mathcal{C}_{q,c,d}^{(n)}$ satisfies $A_{2,c,d}^{(n)}(l)=A_{2,c,d}^{(n)}(n-l)$ for $0\leq l\leq n$. This property simply follows from the fact that for even $d$ the all-one vector is a codeword of $\mathcal{C}_{2,c,d}^{(n)}$. In particular we have the following:

\begin{remark}\label{re:SymmetricProperty}
For even $d\ge 2$,
\begin{equation}\label{eq:SymmetricProperty1}
E{\left[A_{2,c,d}^{(n)}(l)\right]} = E{\left[A_{2,c,d}^{(n)}(n-l)\right]}
\end{equation}
\begin{equation}\label{eq:SymmetricProperty2}
\omega_{2,c,d}(x) = \omega_{2,c,d}(1-x).
\end{equation}
\end{remark}

We close this section with a theorem on the asymptotic behavior of the average weight distribution for the small-weight case.

\begin{theorem}\label{th:LDPCWeigthDistributionL0}
For $d\ge 3$ and constant weight $l\ge 1$,
\[
E{\left[A_{q,c,d}^{(n)}(l)\right]}
= \left\{\begin{array}{ll}
0 &\mbox{$c = 1$ and $l = 1$} \\
0 &\mbox{$q=2$ and $cl$ is odd} \\
\order\left(n^{-\ceil{(c-2)l/2}}\right) &\mbox{otherwise.}
\end{array}\right.
\]
\end{theorem}

\begin{IEEEproof}
The trick of the proof is to find a precise approximation of $\coef(g_{q,d}^{(cn/d)}(x), x^{cl})$ in \eqref{eq:LDPCWeightDistribution} and to prove it by induction. For convenience, we define
\[
A(n,m) \eqdef \coef{\left(g_{q,d}^{(n)}(x), x^m\right)}.
\]
After some algebraic manipulations, we have
\[
g_{q,d}^{(n)}(x) = \left[ \sum_{i=0}^d {d \choose i} B(i) x^i \right]^n
\]
where
\[
B(i) = \frac{(q-1)^i + (-1)^i(q-1)}{q}.
\]
Then it is observed that
\begin{IEEEeqnarray*}{rCl}
A(n+1, m) &= &\sum_{i=0}^{\min\{m,d\}} {d \choose i} A(n,m-i) B(i) \\
&= &A(n, m) + \sum_{i=2}^{\min\{m,d\}} {d \choose i} A(n,m-i) B(i).
\end{IEEEeqnarray*}
Hence we have
\begin{IEEEeqnarray*}{rCl}
A(n,0) &= &A(1, 0) = 1 \\
A(n,1) &= &A(1,1) = 0 \\
A(n,2) &= &A(n-1,2) + {d \choose 2} A(n-1,0) B(2) \\
&= &A(n-1,2) + \frac{d(d-1)(q-1)}{2} \\
&= &\order\left(n^{\floor{\frac{2}{2}}}\right) \\
A(n,3) &= &A(n-1, 3) + {d \choose 2} A(n-1,1) B(2) \\
& &\breakop{+} {d \choose 3} A(n-1,0) B(3) \\
&= &A(n-1, 3) + \frac{d(d-1)(d-2)(q-1)(q-2)}{6} \\
&= &\left\{\begin{array}{ll}
0 &\mbox{$q=2$} \\
\order\left(n^{\floor{\frac{3}{2}}}\right) &\mbox{otherwise.}
\end{array}\right.
\end{IEEEeqnarray*}
We shall show by induction on $m$ that
\begin{equation}
A(n,m) = \left\{\begin{array}{ll}
0 &\mbox{$q=2$ and $m$ is odd} \\
\order\left(n^{\floor{\frac{m}{2}}}\right) &\mbox{otherwise.}
\end{array}\right.\label{eq:Eq1InProofOfLDPCWDL0Thm}
\end{equation}
for all constant $m \ge 2$. Here, we only prove the general case of $q > 2$. The case of $q=2$ can be proved by a similar argument with the fact $B(i)=[1+(-1)^i]/2$. Suppose that \eqref{eq:Eq1InProofOfLDPCWDL0Thm} holds for $2 \le m \le k$ with $k \ge 3$, then for $m = k+1$,
\begin{IEEEeqnarray*}{rCl}
A(n,k+1)
&= &A(n-1,k+1) \\
& &\breakop{+} \sum_{i=2}^{\min\{k+1,d\}} {d \choose i} A(n-1,k-i+1) B(i) \\
&= &A(n-1,k+1) + \order\left((n-1)^{\floor{(k-1)/2}}\right)
\end{IEEEeqnarray*}
This asymptotic behavior implies that there exits a positive integer $n_0$ such that for $n>n_0$,
\begin{IEEEeqnarray*}{rCl}
A(n,k+1) &= &A(n_0, k+1) + \order\left(\sum_{i=n_0}^{n-1} i^{\floor{(k-1)/2}}\right) \\
&= &\order\left(n^{\floor{(k+1)/2}}\right).
\end{IEEEeqnarray*}
Thus \eqref{eq:Eq1InProofOfLDPCWDL0Thm} holds for all $m \ge 2$.

Finally, it follows from Theorem~\ref{th:LDPCWeightDistribution} and \eqref{eq:Eq1InProofOfLDPCWDL0Thm} that 
\begin{IEEEeqnarray*}{rCl}
E{\left[A_{q,c,d}^{(n)}(l)\right]}
&= &\frac{{n \choose l} A(cn/d, cl)}{{cn \choose cl} (q-1)^{(c-1)l}} \\
&= &\left\{\begin{array}{ll}
0 &\mbox{$c = 1$ and $l = 1$} \\
0 &\mbox{$q=2$ and $cl$ is odd} \\
\order\left(n^{-\ceil{(c-2)l/2}}\right) &\mbox{otherwise}
\end{array}\right.
\end{IEEEeqnarray*}
as desired.
\end{IEEEproof}

\begin{remark}
The first and second cases of Theorem~\ref{th:LDPCWeigthDistributionL0} have the following alternative proofs: If $c=1$ then the random code  $\mathcal{C}_{q,c,d}^{(n)}$, as the kernel of the reduced mapping $\chkcode{q,d,n/d} \circ \rmcode{q,n} \circ \Sigma_{q,n}$, has the same weight distribution as the kernel of $\chkcode{q,d,n/d}$. In particular, $\mathcal{C}_{q,c,d}^{(n)}$ has no words of weight $1$. If $c$ is odd then every column of the parity-check matrix of $\mathcal{C}_{2,c,d}^{(n)}$ (i.e. the transformation matrix of $\ldcode{2,c,d,n}$) has odd weight. This implies that the all-one vector is in the dual code of $\mathcal{C}_{2,c,d}^{(n)}$ and hence that all codewords have even weight.
\end{remark}

\section{Properties of the Function $\delta_{q,d}(x)$}\label{sec:Function1}

As an important step towards understanding the function $\omega_{q,c,d}(x)$, we analyze in this section the function $\delta_{q,d}(x)$ defined by \eqref{eq:DefinitionOfDelta}. The proofs of lemmas in this section are presented in Appendix~\ref{app:Proof1}.

In the sequel, we shall frequently use the following substitution to facilitate the analysis:
\begin{equation}
z \eqdef 1-\frac{qx}{q-1}, \quad \hat{z} \eqdef 1-\frac{q\hat{x}}{q-1}. \label{eq:TransfromZX}
\end{equation}
Note that this transform is bijective and strictly decreasing, so we have
\begin{equation}
x = \frac{(q-1)(1-z)}{q}, \quad \hat{x} = \frac{(q-1)(1-\hat{z})}{q} \label{eq:TransfromXZ}
\end{equation}
and $z,\hat{z} \in [-1/(q-1),1]$ as $x,\hat{x} \in [0,1]$.

Our first goal is to study the zeros of the partial derivative of $\delta_{q,d}(x,\hat{x})$ with respect to $\hat{x}$.

\begin{lemma}\label{le:PDOfDelta2WithXHat}
For the function $\delta_{q,d}(x,\hat{x})$ defined by \eqref{eq:DefinitionOfDelta2},
\begin{IEEEeqnarray}{rCl}
\frac{\partial\delta_{q,d}(x,\hat{x})}{\partial\hat{x}}
&= &d\frac{\partial D(x\|\hat{x})}{\partial\hat{x}} + \frac{d\rho_{q,d}(\hat{x})}{d\hat{x}} \label{eq:PDOfDelta2WithXHat1} \\
&= &-\frac{qd(\zeta_{q,d}(\hat{z})-z)}{(1-\hat{z})[1+(q-1)\hat{z}]}
\end{IEEEeqnarray}
where
\begin{equation}
\zeta_{q,d}(\hat{z}) \eqdef \frac{\hat{z}+\hat{z}^{d-1}+(q-2)\hat{z}^d}{1+(q-1)\hat{z}^d}. \label{eq:DefinitionOfZeta}
\end{equation}
\end{lemma}

Lemma~\ref{le:PDOfDelta2WithXHat} shows that the zeros of $\partial\delta_{q,d}(x,\hat{x})/\partial\hat{x}$ are determined by the equation $\zeta_{q,d}(\hat{z})-z=0$. We therefore proceed to analyze the function $\zeta_{q,d}(\hat{z})$. The next three lemmas give the properties of $\zeta_{q,d}(\hat{z})$.

\begin{lemma}\label{le:Property1OfZeta}
For $q\ge 2$ and $d\ge 3$, the function $\zeta_{q,d}(\hat{z})$ is continuously differentiable on $[-1/(q-1), 1]$ and its derivative is positive on $(-1/(q-1),1)$.
\end{lemma}

\begin{lemma}\label{le:Property2OfZeta}
For $q\ge 2$ and $d \ge 1$,
\begin{equation}
\zeta_{q,d}(z)-z = \frac{z^{d-1}(1-z)[1+(q-1)z]}{1+(q-1)z^d} \label{eq:Property2OfXiEq1}
\end{equation}
\begin{subnumcases}{\zeta_{q,d}{\left(-\frac{1}{q-1}\right)} =}
\textstyle\frac{2}{d}-1 &$q=2$ and $d$ is odd \label{eq:Property2OfXiEq2a} \\
\textstyle-\frac{1}{q-1} &otherwise \label{eq:Property2OfXiEq2b}
\end{subnumcases}
\begin{equation}
\zeta_{q,d}(0)=0 \label{eq:Property2OfXiEq3}
\end{equation}
\begin{equation}
\zeta_{q,d}(1)=1. \label{eq:Property2OfXiEq4}
\end{equation}
\end{lemma}

\begin{lemma}\label{le:Property3OfZeta}
Let
\begin{subnumcases}{z_1 \eqdef}
\frac{2}{d}-1 &$q=2$ and $d$ is odd \\
-\frac{1}{q-1} &otherwise.
\end{subnumcases}
The equation $\zeta_{q,d}(\hat{z})-z=0$ has a unique solution $\hat{z}_1=\hat{z}_1(z)$ in $[-1/(q-1),1]$ for each $z\in [z_1,1]$ and has no solution in $[-1/(q-1),1]$ for $z < z_1$. The solution $\hat{z}_1(z)$ is continuous on $[z_1, 1]$ and is continuously differentiable on $(z_1, 1)$; its derivative is positive on $(z_1, 1)$. Moreover, $\hat{z}_1(z) \in I_{q,d}'(z)$, where
\begin{numcases}{I_{q,d}'(z)\eqdef}
\textstyle \{-\frac{1}{q-1}\} &$z=z_1$ \IEEEnonumber\\
\textstyle (-\frac{1}{q-1},z) &$z\in(z_1,0)$ and $d$ is odd \IEEEnonumber\\
(z,0) &$z\in(z_1,0)$ and $d$ is even \IEEEnonumber\\
\{0\} &$z=0$ \IEEEnonumber\\
(0,z) &$z\in(0,1)$ \IEEEnonumber\\
\{1\} &$z=1$. \IEEEnonumber
\end{numcases}
\end{lemma}

Equipped with Lemmas~\ref{le:PDOfDelta2WithXHat}--\ref{le:Property3OfZeta}, we are now in a position to analyze the function $\delta_{q,d}(x)$.

\begin{theorem}\label{th:PropertyOfDelta}
Let $q \ge 2$, $d \ge 3$, and
\begin{subnumcases}{x_1 \eqdef\label{eq:DefinitionOfX1}}
1-\frac{1}{d} &$q=2$ and $d$ is odd \\
1 &otherwise.
\end{subnumcases}
For the function $\delta_{q,d}(x)$ defined by \eqref{eq:DefinitionOfDelta}, we have
\begin{subnumcases}{\delta_{q,d}(x)=}
\ln q &$x=0$ \label{eq:PropertyEq1aOfDelta} \\
\rho_{q,d}(1) &$x=1$ \label{eq:PropertyEq1bOfDelta} \\
-\infty &$x\in(1-\frac{1}{d},1)$, $q=2$, \IEEEnonumber\\
&and $d$ is odd \label{eq:PropertyEq1cOfDelta} \\
\ln(2d) - dH_2{\left(\frac{1}{d}\right)} &$x=1-\frac{1}{d}$, $q=2$, \IEEEnonumber\\
&and $d$ is odd \label{eq:PropertyEq1dOfDelta} \\
\delta_{q,d}(x,\hat{x}_1) &$x \in (0, x_1)$ \label{eq:PropertyEq1eOfDelta}
\end{subnumcases}
where $\rho_{q,d}(x)$ is defined by \eqref{eq:DefinitionOfRho} and $\hat{x}_1=\hat{x}_1(x)$ is the unique root in $(0,1)$ of the equation
\begin{equation}\label{eq:DefinitionOfXHat1}
\frac{\partial\delta_{q,d}(x,\hat{x})}{\partial\hat{x}}=0
\end{equation}
solved for $\hat{x}$ as a function of $x$. The function $\hat{x}_1(x)$ is continuously differentiable on $(0, x_1)$ and its derivative is positive on $(0, x_1)$. Moreover, $\lim_{x\to 0^+} \hat{x}_1(x) = 0$, $\lim_{x\to x_1^-} \hat{x}_1(x) = 1$, and $\hat{x}_1(x) \in I_{q,d}(x)$, where
\begin{numcases}{I_{q,d}(x) \eqdef}
\textstyle(x,1-\frac{1}{q}) &$x\in (0,1-\frac{1}{q})$ \IEEEnonumber\\
\textstyle\{1-\frac{1}{q}\} &$x=1-\frac{1}{q}$ \IEEEnonumber\\
(x, 1) &$x\in(1-\frac{1}{q}, x_1)$ and $d$ is odd \IEEEnonumber\\
\textstyle(1-\frac{1}{q}, x) &$x\in(1-\frac{1}{q}, x_1)$ and $d$ is even. \IEEEnonumber
\end{numcases}
The function $\delta_{q,d}(x)$ is continuous on $[0, x_1]$ and is continuously differentiable on $(0, x_1)$, in which case,
\begin{equation}
\frac{d\delta_{q,d}(x)}{dx}
= d\ln \frac{x(1-\hat{x}_1)}{\hat{x}_1(1-x)}. \label{eq:PropertyEq2OfDelta}
\end{equation}
\end{theorem}

\begin{IEEEproof}
At first, Lemmas~\ref{le:PDOfDelta2WithXHat}, \ref{le:Property1OfZeta}, and \ref{le:Property2OfZeta} show that
\[
\frac{\partial \delta_{q,d}(0, \hat{x})}{\partial\hat{x}} > 0 \qquad \forall \hat{x}\in(0,1)
\]
and
\[
\frac{\partial \delta_{q,d}(1, \hat{x})}{\partial\hat{x}} < 0 \qquad \forall \hat{x}\in(0,1).
\]
Therefore we have
\[
\delta_{q,d}(0) = \lim_{\hat{x} \to 0^+} \delta_{q,d}(0, \hat{x}) = \rho_{q,d}(0)
\]
and
\[
\delta_{q,d}(1) = \lim_{\hat{x} \to 1^-} \delta_{q,d}(1, \hat{x}) = \rho_{q,d}(1).
\]
This concludes \eqref{eq:PropertyEq1aOfDelta} and \eqref{eq:PropertyEq1bOfDelta}.

A similar argument also shows that for odd $d$
\[
\frac{\partial \delta_{2,d}(x, \hat{x})}{\partial\hat{x}} < 0 \qquad \forall x\in\left[1-\frac{1}{d},1\right), \hat{x}\in(0,1)
\]
so that
\begin{IEEEeqnarray*}{rCl}
\delta_{2,d}(x)
&= & \lim_{\hat{x} \to 1^-} \delta_{2,d}(x, \hat{x}) \\
&= & -dH_2(x) + \lim_{\hat{x} \to 1^-} \ln \frac{1+(1-2\hat{x})^d}{(1-\hat{x})^{d(1-x)}} \\
&= & -dH_2(x) + \ln \lim_{\hat{x} \to 1^-} \frac{2(1-2\hat{x})^{d-1}}{(1-x)(1-\hat{x})^{d(1-x)-1}} \\
&= & -dH_2(x) + \ln \frac{2}{(1-x) \lim_{\hat{x} \to 1^-} (1-\hat{x})^{d-1-dx}}
\end{IEEEeqnarray*}
which yields \eqref{eq:PropertyEq1cOfDelta} and \eqref{eq:PropertyEq1dOfDelta}.

For $x\in(0, x_1)$, Lemma~\ref{le:Property3OfZeta} shows that there is a unique $\hat{z}_1=\hat{z}_1(z) \in (-1/(q-1),1)$ such that $\zeta_{q,d}(\hat{z}_1) = z =1-qx/(q-1)$. Let $\hat{x}_1 = (q-1)(1-\hat{z}_1)/q$, which is essentially a function of $x$. Then it follows from Lemma~\ref{le:PDOfDelta2WithXHat} and \ref{le:Property1OfZeta} that
\[
\frac{\partial \delta_{q,d}(x,\hat{x})}{\partial\hat{x}} < 0 \qquad \forall \hat{x}\in (0,\hat{x}_1)
\]
and
\[
\frac{\partial \delta_{q,d}(x,\hat{x})}{\partial\hat{x}} > 0 \qquad \forall \hat{x}\in (\hat{x}_1, 1).
\]
Therefore, $\delta_{q,d}(x) = \delta_{q,d}(x,\hat{x}_1)$, which concludes \eqref{eq:PropertyEq1eOfDelta}. Furthermore, Lemma~\ref{le:Property3OfZeta} shows that $\hat{x}_1(x)$ is continuously differentiable on $(0, x_1)$ and its derivative is positive on $(0, x_1)$. It also shows that $\lim_{x\to 0^+} \hat{x}_1(x) = 0$ and $\lim_{x\to x_1^-} \hat{x}_1(x) = 1$, and that $\hat{x}_1(x) \in I_{q,d}(x)$.

Based on the above analysis, it is clear that $\delta_{q,d}(x)$ is continuously differentiable on $(0, x_1)$. Furthermore, equation \eqref{eq:PropertyEq1eOfDelta} combined with Lemma~\ref{le:Derivatives} gives \eqref{eq:PropertyEq2OfDelta}.

Finally, let us show that $\delta_{q,d}(x)$ is continuous at the endpoints of the interval. Note that $\delta_{q,d}(x)$ is the infimum of a collection of continuous functions, so it is upper semi-continuous. Then it suffices to show that $\lim_{x\to 0^+} \delta_{q,d}(x) \ge \delta_{q,d}(0)$ and $\lim_{x\to x_1^-} \delta_{q,d}(x) \ge \delta_{q,d}(x_1)$. Recall that $\lim_{x\to 0^+} \hat{x}_1(x) = 0$ and $\lim_{x\to x_1^-} \hat{x}_1(x) = 1$, so we have
\[
\lim_{x\to 0^+} \delta_{q,d}(x) \ge \lim_{x\to 0^+} \rho_{q,d}(\hat{x}_1(x)) = \ln q
\]
\[
\lim_{x\to x_1^-} \delta_{q,d}(x) \ge \lim_{x\to x_1^-} \rho_{q,d}(\hat{x}_1(x)) = \rho_{q,d}(1)
\]
and
\begin{IEEEeqnarray*}{rCl}
\lim_{x\to x_1^-} \delta_{2,d}(x)
&\ge &\lim_{x \to x_1^-} \left[-dH_2(x) + \ln \frac{1+(1-2\hat{x}_1(x))^d}{1-\hat{x}_1(x)}\right] \\
&= &\ln(2d) - dH_2{\left(\frac{1}{d}\right)}
\end{IEEEeqnarray*}
for odd $d$. The proof is complete.
\end{IEEEproof}

\begin{figure*}[htbp]
\centering
\includegraphics{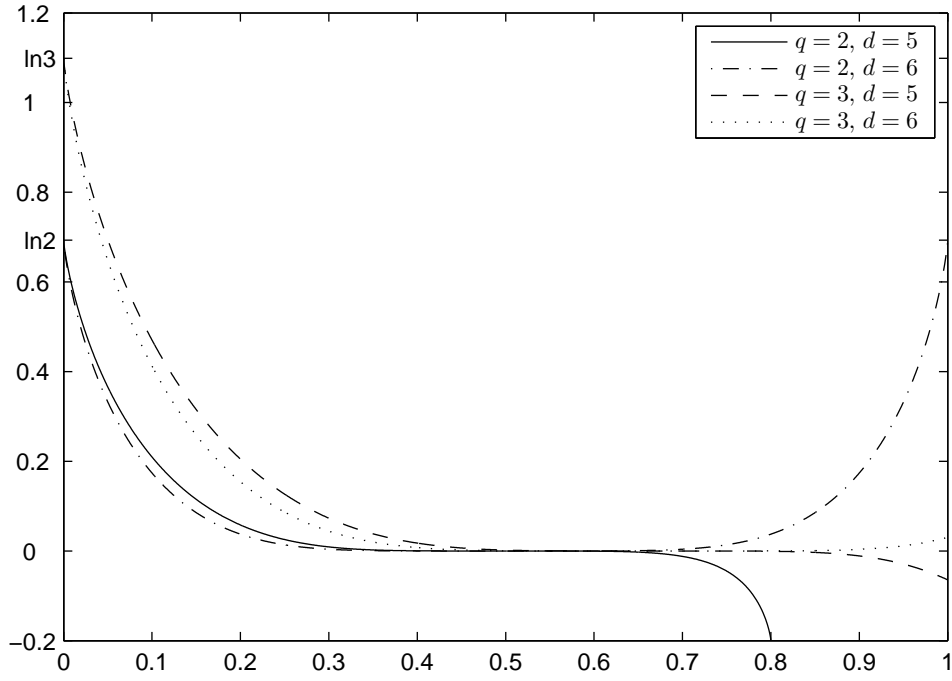}
\caption{The graphs of $\delta_{q,d}(x)$ for $(q,d)=(2,5)$, $(q,d)=(2,6)$,  $(q,d)=(3,5)$, and $(q,d)=(3,6)$.}
\label{fig:Delta1}
\end{figure*}

In Fig.~\ref{fig:Delta1} we give an illustration of the graphs of $\delta_{q,d}(x)$ for $(q,d)=(2,5)$, $(q,d)=(2,6)$,  $(q,d)=(3,5)$, and $(q,d)=(3,6)$.

\section{Properties of the Function $\omega_{q,c,d}(x)$}\label{sec:Function2}

In this section, we proceed to analyze the properties of the function $\omega_{q,c,d}(x)$ defined by \eqref{eq:DefinitionOfOmega}. Since LDPC codes are trivial when $c > d$, we shall sometimes assume $c \le d$ to exclude trivial cases. The proofs of lemmas in this section are presented in Appendix~\ref{app:Proof2}.

At first, we calculate the value of $\omega_{q,c,d}(x)$ at some special points.

\begin{lemma}\label{le:OmegaAtSpecialPoints}
Let $q \ge 2$, $c \ge 1$, and $d \ge 3$.
\begin{equation}
\omega_{q,c,d}(0) = 0.
\end{equation}
\begin{equation}
\omega_{q,c,d}{\left(1-\frac{1}{q}\right)} = \left(1-\frac{c}{d}\right) \ln q.
\end{equation}
\begin{equation}
\omega_{q,c,d}(1) = \ln(q-1)+\frac{c}{d}\rho_{q,d}(1)-\frac{c}{d} \ln q.
\end{equation}
If $q=2$ and $d$ is odd then
\begin{equation}
\omega_{q,c,d}{\left(1-\frac{1}{d}\right)} = \left(1-c\right) H_2{\left(\frac{1}{d}\right)} + \frac{c}{d} \ln d
\end{equation}
and
\begin{equation}
\omega_{q,c,d}(x) = -\infty \qquad \forall x \in \left(1-\frac{1}{d}, 1\right).
\end{equation}
\end{lemma}

Lemma~\ref{le:OmegaAtSpecialPoints} is an easy consequence of Theorem~\ref{th:PropertyOfDelta}, so its proof is left to the reader. Next, let us calculate the first-order derivative of $\omega_{q,c,d}(x)$.

\begin{lemma}\label{le:D1OfOmega}
For the function $\omega_{q,c,d}(x)$ defined by \eqref{eq:DefinitionOfOmega} with $q \ge 2$, $c \ge 1$, and $d \ge 3$, if $x$ belongs to the case \eqref{eq:PropertyEq1eOfDelta} then
\begin{equation}\label{eq:D1OfOmega1}
\frac{d\omega_{q,c,d}(x)}{dx}
= \ln \left[\left(\frac{x}{1-x}\right)^{c-1} \left(\frac{1-\hat{x}_1}{\hat{x}_1}\right)^{c}\right] + \ln(q-1)
\end{equation}
which can be further expressed as
\begin{equation}\label{eq:D1OfOmega2}
\frac{d\omega_{q,c,d}(x)}{dx}
= \ln \left\{\frac{1+(q-1)\hat{z}_1}{1-\hat{z}_1} \left[\frac{1-\hat{z}_1^{d-1}}{1+(q-1)\hat{z}_1^{d-1}}\right]^{c-1}\right\}
\end{equation}
where $\hat{x}_1$ is defined by \eqref{eq:DefinitionOfXHat1} and $\hat{z}_1 = 1 - q\hat{x}_1/(q-1)$.
\end{lemma}

The next lemma gives the value of $d\omega_{q,c,d}(x)/dx$ at some special points.

\begin{lemma}\label{le:D1OfOmegaAtSpecialPoints}
Let $q \ge 2$, $d \ge 3$, and $x_1$ be defined by \eqref{eq:DefinitionOfX1}.
\begin{subnumcases}{\lim_{x\to 0^+} \frac{d\omega_{q,c,d}(x)}{dx} =\label{eq:D1OfOmegaAtSpecialPoints1}}
\infty &$c=1$ \label{eq:D1OfOmegaAtSpecialPoints1a}\\
\ln(d-1) &$c=2$ \label{eq:D1OfOmegaAtSpecialPoints1b}\\
-\infty &$c\ge 3$. \label{eq:D1OfOmegaAtSpecialPoints1c}
\end{subnumcases}
\begin{equation}\label{eq:D1OfOmegaAtSpecialPoints2}
\left.\frac{d\omega_{q,c,d}(x)}{dx}\right|_{x=1-\frac{1}{q}} = 0.
\end{equation}
If $q=2$ and $d$ is even then
\begin{subnumcases}{\lim_{x\to 1^-} \frac{d\omega_{q,c,d}(x)}{dx} =\label{eq:D1OfOmegaAtSpecialPoints3}}
-\infty &$c=1$ \\
-\ln(d-1) &$c=2$ \\
\infty &$c\ge 3.$
\end{subnumcases}
If $q \ne 2$ or $d$ is odd then
\begin{equation}\label{eq:D1OfOmegaAtSpecialPoints4}
\lim_{x\to x_1^-} \frac{d\omega_{q,c,d}(x)}{dx} = -\infty.
\end{equation}
\end{lemma}

To have more insights into $\omega_{q,c,d}(x)$, we proceed to analyze the second-order derivative of $\omega_{q,c,d}(x)$. Since
\begin{equation}\label{eq:D2OfOmega}
\frac{d^2\omega_{q,c,d}(x)}{dx^2}
= \frac{d}{d\hat{z}_1} {\left(\frac{d\omega_{q,c,d}(x)}{dx}\right)} \cdot \frac{d\hat{z}_1}{dx}
\end{equation}
and we note that
\[
\frac{d\hat{z}_1}{dx} = -\frac{q}{q-1} \frac{d\hat{x}_1}{dx}
\]
is negative on $(0, x_1)$, our task is now to calculate the derivative $d(d\omega_{q,c,d}(x)/dx)/d\hat{z}_1$.

\begin{lemma}\label{le:D2ZOfOmega}
For the function $\omega_{q,c,d}(x)$ defined by \eqref{eq:DefinitionOfOmega} with $q \ge 2$, $c \ge 1$, and $d \ge 3$, if $x$ belongs to the case \eqref{eq:PropertyEq1eOfDelta} then
\begin{IEEEeqnarray}{l}
\frac{d}{d\hat{z}_1} {\left(\frac{d\omega_{q,c,d}(x)}{dx}\right)} \IEEEnonumber\\
\findent = \frac{q\xi_{q,c,d}(\hat{z}_1)}{(1-\hat{z}_1^{d-1})[1+(q-1)\hat{z}_1][1+(q-1)\hat{z}_1^{d-1}]} \label{eq:D2ZOfOmega1}
\end{IEEEeqnarray}
where
\begin{IEEEeqnarray}{rCl}
\xi_{q,c,d}(\hat{z})
&= &\sum_{i=0}^{d-3} \hat{z}^i - [(c-1)(d-1)-1] \hat{z}^{d-2} \IEEEnonumber\\
& &\breakop{-} (q-1)[(c-1)(d-1)-1] \hat{z}^{d-1} \IEEEnonumber\\
& &\breakop{+} (q-1) \sum_{i=d}^{2d-3} \hat{z}^i. \label{eq:DefinitionOfXi}
\end{IEEEeqnarray}
When $c=1$, equation \eqref{eq:D2ZOfOmega1} reduces to
\begin{equation}\label{eq:D2ZOfOmega2}
\frac{d}{d\hat{z}_1} {\left(\frac{d\omega_{2,c,d}(x)}{dx}\right)}
= \frac{q}{(1-\hat{z}_1)[1+(q-1)\hat{z}_1]}.
\end{equation}
\end{lemma}

We go on to analyze the function $\xi_{q,c,d}(\hat{z})$ for $q \ge 2$, $c \ge 2$, and $d \ge \max\{c,3\}$.

\begin{lemma}\label{le:PropertyOfXi}
For $d \ge 3$, the function $\xi_{2,2,d}(\hat{z})$ is positive on $(-1, 1)$. For $q\ge 3$ and $d \ge 3$, the function $\xi_{q,2,d}(\hat{z})$ has a positive zero $\hat{z}_2$ in $(-1/(q-1), 1)$, and $\xi_{q,2,d}(\hat{z})$ is positive on $(-1/(q-1), \hat{z}_2)$ and negative on $(\hat{z}_2, 1)$.

For $d \ge c \ge 3$ with $d$ even, the function $\xi_{2,c,d}(\hat{z})$ has one zero $\hat{z}_2$ in $(0, 1)$ and the other zero $\hat{z}_2'$ in $(-1, 0)$, and $\xi_{2,c,d}(\hat{z})$ is positive on $(\hat{z}_2', \hat{z}_2)$ and negative on $(-1, \hat{z}_2') \cup (\hat{z}_2, 1)$.

For $q \ge 2$ and $d \ge c \ge 3$ with $q \ne 2$ or $d$ odd, the function $\xi_{q,c,d}(\hat{z})$ has a positive zero $\hat{z}_2$ in $(-1/(q-1), 1)$, and $\xi_{q,c,d}(\hat{z})$ is positive on $(-1/(q-1), \hat{z}_2)$ and negative on $(\hat{z}_2, 1)$.
\end{lemma}

We are now ready to give the qualitative properties of $\omega_{q,c,d}(x)$.

\begin{theorem}\label{th:PropertyOfOmega}
Let $q \ge 2$, $c \ge 1$, $d \ge \max\{c,3\}$, and $x_1$ be defined by \eqref{eq:DefinitionOfX1}. The function $\omega_{q,c,d}(x)$ defined by \eqref{eq:DefinitionOfOmega} is continuous on $[0, x_1]$ and is twice differentiable on $(0, x_1)$.

If $c=1$, then $\omega_{q,c,d}(x)$ is concave on $(0, x_1)$, and it is strictly increasing on $(0, 1-1/q)$ and strictly decreasing on $(1-1/q, x_1)$.

If $c=2$, then $\omega_{q,c,d}(x)$ is strictly increasing on $(0, 1-1/q)$ and strictly decreasing on $(1-1/q, x_1)$. Moreover, if $q=2$, it is concave on $(0, x_1)$; otherwise, it is convex on $(0, x_2)$ and concave on $(x_2, 1)$, where $x_2 \in (0, 1-1/q)$.

If $c\ge 3$, $q=2$, and $d$ is even, then $\omega_{q,c,d}(x)$ is symmetric about the axis $x=\frac{1}{2}$. It is convex on $(0, x_2)$ and concave on $(x_2, \frac{1}{2})$ for some $x_2 \in (0, \frac{1}{2})$; it is strictly decreasing on $(0,x_3)$ and strictly increasing on $(x_3, \frac{1}{2})$, where $x_3 \in (0, x_2)$; consequently, it has a unique zero $x_0$ in $(0, \frac{1}{2}]$, where $x_0 \in (x_3, \frac{1}{2}]$, and it is negative on $(0, x_0)$ and positive on $(x_0, \frac{1}{2})$.

For other cases, the function $\omega_{q,c,d}(x)$ is convex on $(0,x_2)$ and concave on $(x_2, x_1)$, where $x_2 \in (0, 1-1/q)$; it is strictly decreasing on $(0,x_3) \cup (1-1/q, x_1)$ and strictly increasing on $(x_3, \frac{1}{2})$, where $x_3 \in (0, x_2)$; consequently, it has a unique zero $x_0$ in $(0, 1-1/q]$, where $x_0 \in (x_3, 1-1/q]$, and it is negative on $(0, x_0)$ and positive on $(x_0, 1-1/q)$.
\end{theorem}

\begin{figure*}[htbp]
\centering
\includegraphics{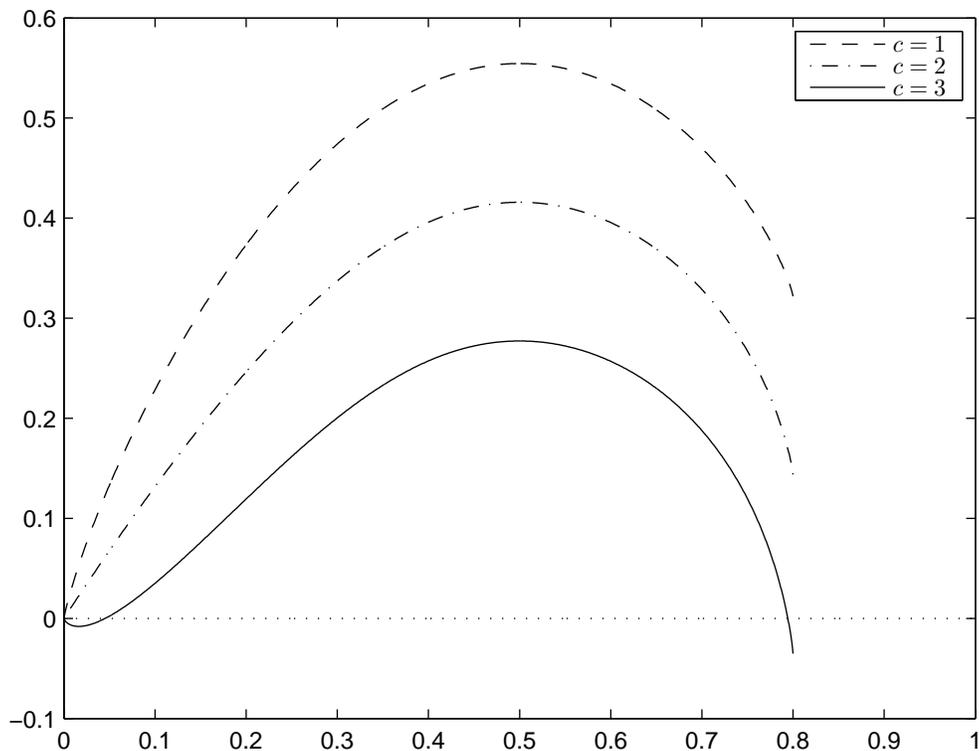}
\caption{The graphs of $\omega_{2,c,5}(x)$ for $c=1$, $c=2$, and $c=3$.}
\label{fig:Omega25}
\end{figure*}

\begin{figure*}[htbp]
\centering
\includegraphics{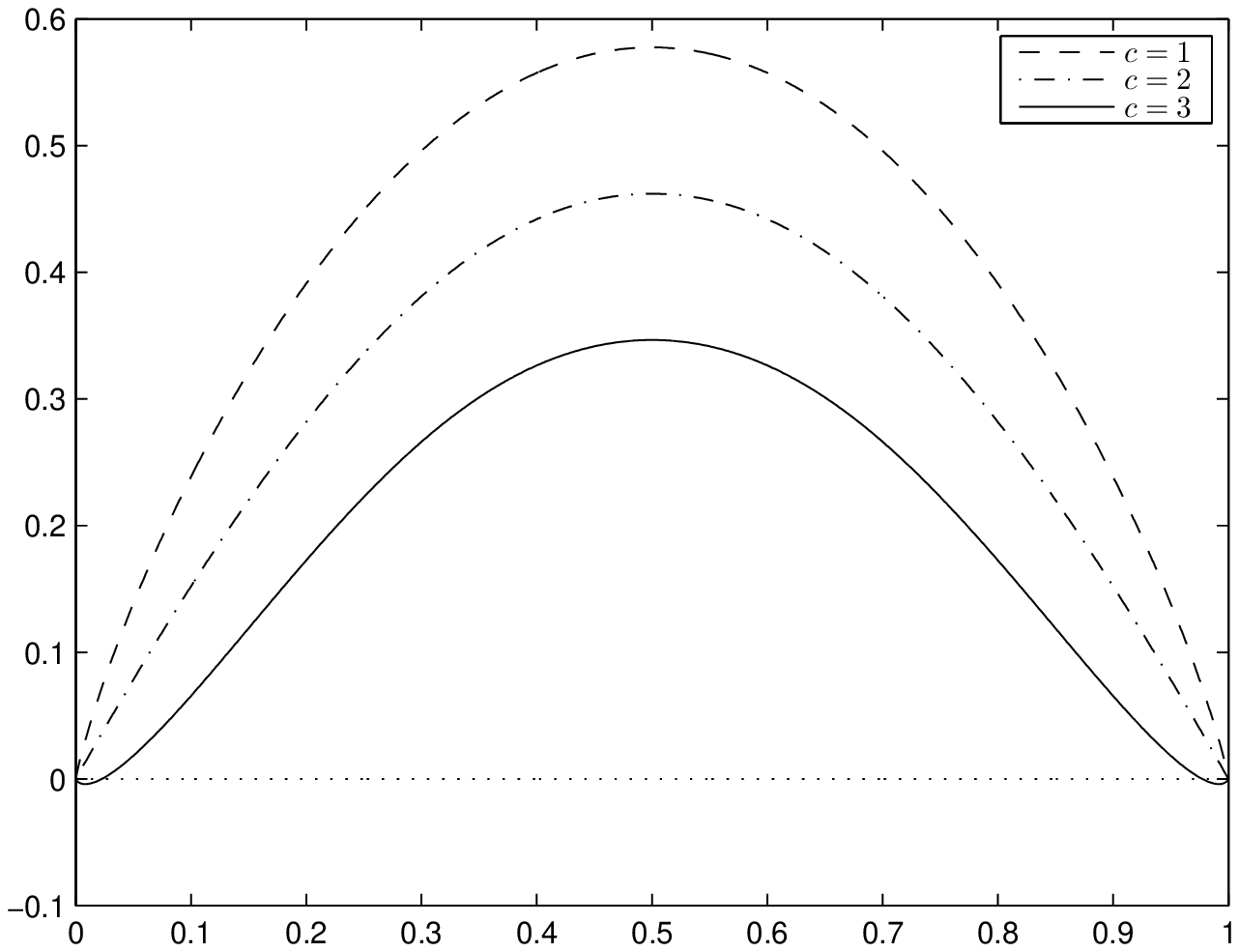}
\caption{The graphs of $\omega_{2,c,6}(x)$ for $c=1$, $c=2$, and $c=3$.}
\label{fig:Omega26}
\end{figure*}

\begin{figure*}[htbp]
\centering
\includegraphics{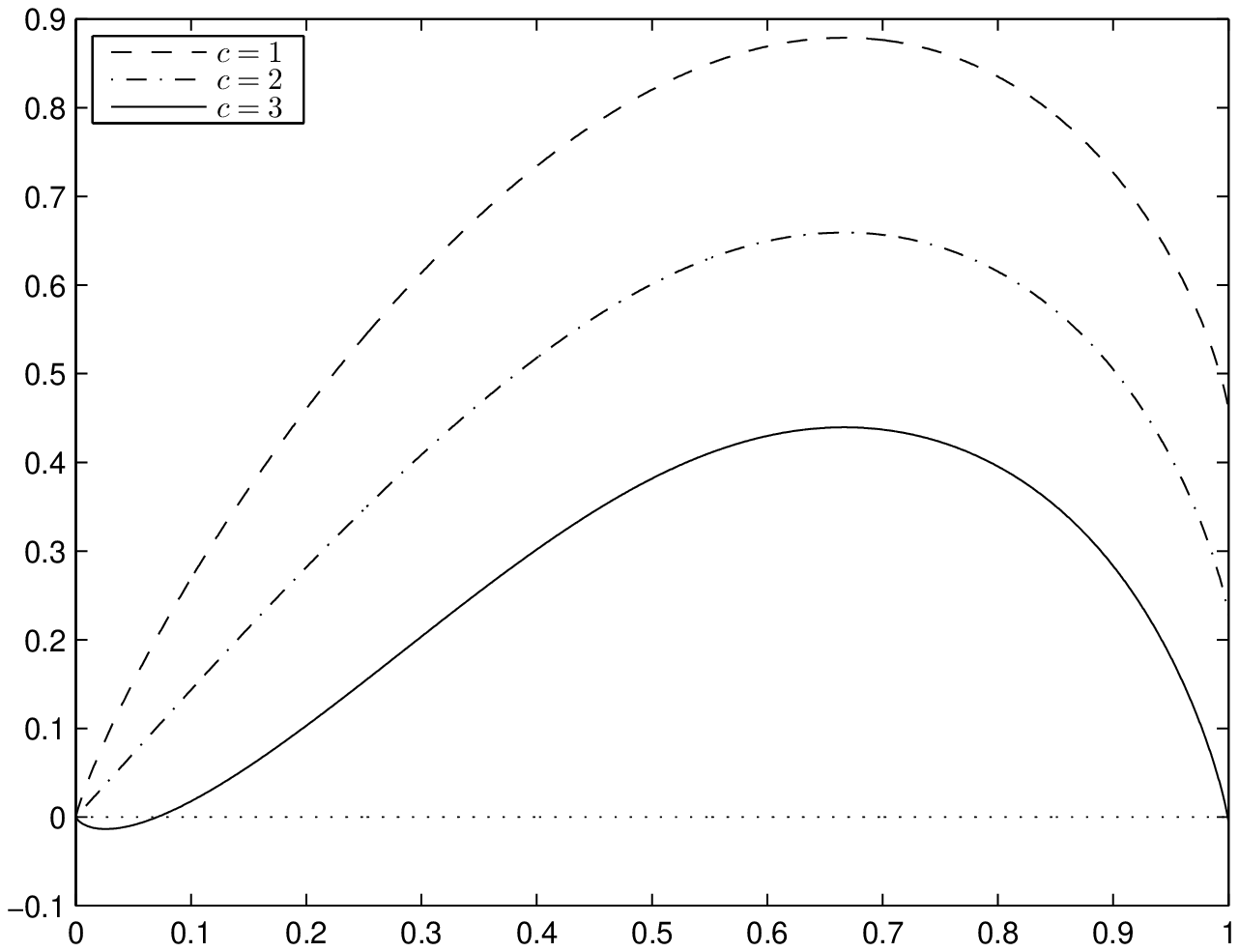}
\caption{The graphs of $\omega_{3,c,5}(x)$ for $c=1$, $c=2$, and $c=3$.}
\label{fig:Omega35}
\end{figure*}

\begin{figure*}[htbp]
\centering
\includegraphics{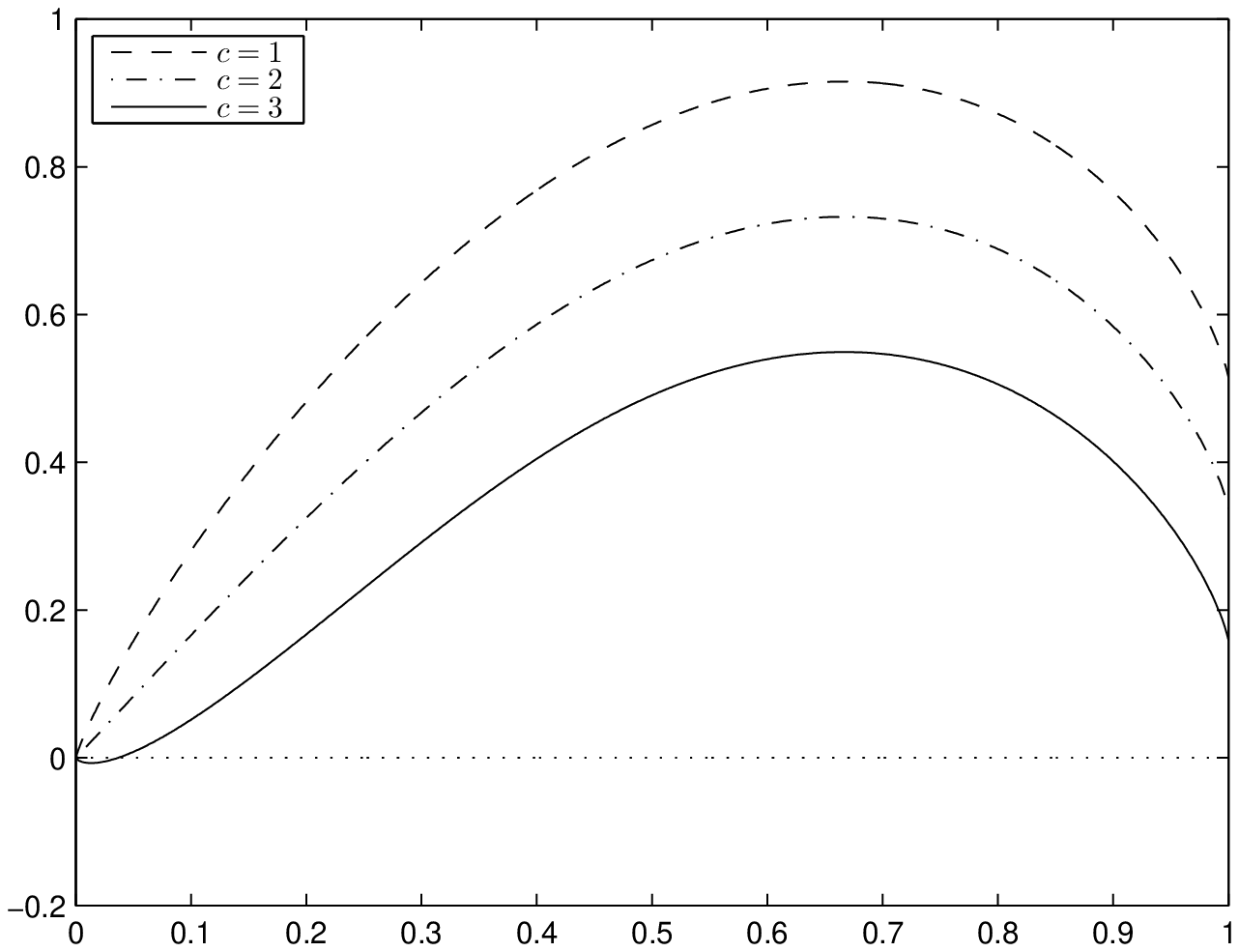}
\caption{The graphs of $\omega_{3,c,6}(x)$ for $c=1$, $c=2$, and $c=3$.}
\label{fig:Omega36}
\end{figure*}

To provide an intuitive illustration of $\omega_{q,c,d}(x)$ in each case, the graphs of $\omega_{q,c,d}(x)$ for typical values of $(q,c,d)$ are plotted in Figs.~\ref{fig:Omega25}--\ref{fig:Omega36}.

\begin{IEEEproof}[Sketch of Proof]
The proof is direct, and it depends on Remark~\ref{re:SymmetricProperty}, Theorem~\ref{th:PropertyOfDelta}, Lemmas~\ref{le:OmegaAtSpecialPoints}--\ref{le:PropertyOfXi}, and identity \eqref{eq:D2OfOmega}. Here, we only give the proof of the last paragraph of statements.

Lemma~\ref{le:PropertyOfXi} and identity \eqref{eq:D2OfOmega} show that $\omega_{q,c,d}$ is convex on $(0, x_2)$ and concave on $(x_2, x_1)$, where $x_2 \in (0, 1-1/q)$. Furthermore, Lemmas~\ref{le:OmegaAtSpecialPoints} and \ref{le:D1OfOmegaAtSpecialPoints} show that
\[
\omega_{q,c,d}(0) = 0, \quad \omega_{q,c,d}{\left(1-\frac{1}{q}\right)} = \left(1-\frac{c}{d}\right) \ln q \ge 0
\]
and
\[
\lim_{x\to 0^+} \frac{d\omega_{q,c,d}(x)}{dx} = -\infty, \quad \left.\frac{d\omega_{q,c,d}(x)}{dx}\right|_{x=1-\frac{1}{q}} = 0.
\]
Therefore, the derivative $d\omega_{q,c,d}(x)/dx$ has a unique zero $x_3$ in $(0, 1-1/q)$, where $x_3 \in (0, x_2)$; it is negative on $(0,x_3) \cup (1-1/q, x_1)$ and positive on $(x_3, 1-1/q)$. In other words, the function $\omega_{q,c,d}(x)$ is strictly decreasing on $(0,x_3) \cup (1-1/q, x_1)$ and strictly increasing on $(x_3, 1-1/q)$. The last statement about the unique zero in $(0, 1-1/q)$ clearly follows.
\end{IEEEproof}

\begin{remark}\label{re:GVBound}
The zero $x_0$ in Theorem~\ref{th:PropertyOfOmega} just corresponds to the normalized minimum distance of LDPC codes, in an average and asymptotic sense. It is in fact a function of $q$, $c$, and $d$, so we denote it by $x_0(q,c,d)$. We note that
\[
\lim_{d \to \infty} \rho_{q,d}(x) = 0 \qquad \forall x \in (0, 1)
\]
and hence for any $r \in (0, 1]$,
\[
\lim_{d \to \infty} x_0(q,\ceil{rd},d) = x_{0,q,r}
\]
where $x_{0,q,r}$ is the solution of $H_q(x) - r\ln q = 0$ in $(0, 1-1/q)$. The detailed proof is left to the reader. Note that $x_{0,q,r}$ as well as the equation $H_q(x) - r\ln q = 0$ is closely related to the so-called asymptotic Gilbert-Varshamov (GV) bound over finite fields \cite[pp. 94--95]{JSCC:Huffman200300}. This implies that regular LDPC codes with large $c$ and $d$ achieve the GV bound.
\end{remark}

\section{Minimum Distance of LDPC Codes}\label{sec:MinimumDistance}

Though we have shown in Remark~\ref{re:GVBound} that regular LDPC code ensembles are asymptotically good, we are more interested in the performance of individual codes of finite length. In this section, we shall investigate the minimum distance of an individual code in a regular LDPC code ensemble. To achieve this goal, we first establish an important inequality.

\begin{theorem}\label{th:InequalityOfOmega}
For $q \ge 2$, $c \ge 1$, $d \ge 2$, and $x \in (0, 1/q^{2})$,
\begin{equation}
\omega_{q,c,d}(x) < \left(\frac{c}{2}-1\right) x \ln x + \kappa_{q,c,d} x
\end{equation}
where
\begin{equation}
\kappa_{q,c,d} \eqdef \ln(q-1) + \frac{c}{2}\ln(d-1) + 3c.
\end{equation}
\end{theorem}

\begin{IEEEproof}
Put
\begin{equation}\label{eq:Eq1InProofOfInequalityOfOmegaLem}
\hat{x} \eqdef \sqrt{\frac{x}{d-1}}.
\end{equation}
Then for any $x \in (0, 1/q^{2})$, $\hat{x} \in (0, 1/q) \subset (0, 1-1/q)$.

According to the definition \eqref{eq:DefinitionOfOmega} of $\omega_{q,c,d}(x)$, we have
\begin{IEEEeqnarray*}{l}
\omega_{q,c,d}(x) \\
\findent \le H_q(x) + \frac{c}{d}(\delta_{q,d}(x,\hat{x}) - \ln q) \\
\findent = H_q(x) + cD(x\|\hat{x}) \\
\findent \nullrel \breakop{+} \frac{c}{d} \ln \left[\frac{1}{q}+\left(1-\frac{1}{q}\right)\left(1-\frac{q\hat{x}}{q-1}\right)^d\right] \\
\findent \levar{(a)} H_q(x) + cD(x\|\hat{x}) + c \left[-\hat{x} + \frac{q(d-1)}{2(q-1)} \hat{x}^2 \right] \\
\findent = -(c-1)H_2(x) + x\ln(q-1) \\
\findent \nullrel \breakop{+} c \left[x\ln\frac{1}{\hat{x}} + (1-x)\ln\frac{1}{1-\hat{x}} - \hat{x} + \frac{q(d-1)}{2(q-1)} \hat{x}^2 \right] \\
\findent < (c-1)x\ln x + x \ln(q-1) \\
\findent \nullrel \breakop{+} c \left[x\ln\frac{1}{\hat{x}} + \frac{\hat{x}}{1-\hat{x}} - \hat{x} + \frac{q(d-1)}{2(q-1)} \hat{x}^2 \right] \\
\findent = (c-1)x\ln x + x \ln(q-1) \\
\findent \nullrel \breakop{+} c \left[x\ln\frac{1}{\hat{x}} + \frac{\hat{x}^2}{1-\hat{x}} + \frac{q(d-1)}{2(q-1)} \hat{x}^2 \right] \\
\findent \eqvar{(b)} (c-1)x\ln x + x \ln(q-1) \\
\findent \nullrel \breakop{+} c \left[\frac{1}{2}x\ln\frac{d-1}{x} + \frac{x}{(d-1)(1-\hat{x})} + \frac{qx}{2(q-1)} \right] \\
\findent \ltvar{(c)} \left(\frac{c}{2}-1\right)x\ln x + \left(\ln(q-1) + \frac{c}{2}\ln(d-1) + 3c \right) x
\end{IEEEeqnarray*}
where (a) follows from Lemma~\ref{le:Inequality2} and $\ln x \le x-1$, (b) from \eqref{eq:Eq1InProofOfInequalityOfOmegaLem}, and (c) follows from $q \ge 2$, $d \ge 2$, and $\hat{x} < 1/q$.
\end{IEEEproof}

Now, let us present the main result on the minimum distance of individual codes in a regular LDPC code ensemble.

\begin{theorem}\label{th:MinimumDistance}
For any code $C \subseteq \field{q}^n$, we denote its minimum distance by $\dmin(C)$. Then for $q\ge 2$, $d \ge c \ge 3$, $l_0 \ge 1$, and $\alpha \in (0, 1-1/q)$,
\begin{IEEEeqnarray}{l}
P{\left\{l_0 \le \dmin(\mathcal{C}_{q,c,d}^{(n)}) \le n\alpha\right\}} \IEEEnonumber \\
\findent \le \order\left(n^{-\ceil{(c-2)(l_0+\Delta)/2}}\right) + \order\left(n^{\frac{3}{2}}e^{n\omega_{q,c,d}(\alpha)}\right) \label{eq:MinimumDistance1}
\end{IEEEeqnarray}
where
\begin{equation}\label{eq:DeltaOfL0}
\Delta \eqdef \left\{\begin{array}{ll}
1 &\mbox{$q=2$ and $cl_0$ is odd} \\
0 &\mbox{otherwise}.
\end{array}\right.
\end{equation}
\end{theorem}

\begin{IEEEproof}
Since the minimum distance of a linear code is the minimum weight of its nonzero codewords, we have
\begin{IEEEeqnarray*}{l}
P{\left\{l_0 \le \dmin(\mathcal{C}_{q,c,d}^{(n)}) \le n\alpha\right\}} \\
\findent \le P{\left\{\bigcup_{l=l_0}^{\floor{n\alpha}} \left\{A_{q,c,d}^{(n)}(l) \ge 1\right\}\right\}} \\
\findent \le \sum_{l=l_0}^{\floor{n\alpha}} P{\left\{A_{q,c,d}^{(n)}(l) \ge 1\right\}} \\
\findent \levar{(a)} \sum_{l=l_0}^{\floor{n\alpha}} E{\left[A_{q,c,d}^{(n)}(l)\right]} \\
\findent \levar{(b)} \sum_{l=l_0}^{l_0+3} \order\left(n^{-\ceil{(c-2)l/2}}\right) + \sum_{l=l_0+4}^{\floor{n\alpha}} \order\left(n^{\frac{1}{2}} e^{n\omega_{q,c,d}(l/n)}\right) \\
\findent \levar{(c)} \order\left(n^{-\ceil{(c-2)l_0/2}}\right) + \order\left(n^{\frac{3}{2}} e^{n\omega_{q,c,d}((l_0+4)/n)}\right) \\
\findent \nullrel \breakop{+} \order\left(n^{\frac{3}{2}} e^{n\omega_{q,c,d}(\alpha)}\right) \\
\findent \levar{(d)} \order\left(n^{-\ceil{(c-2)l_0/2}}\right) + \order\left(n^{\frac{3}{2}} n^{-(c-2)(l_0+4)/2}\right) \\
\findent \nullrel \breakop{+} \order\left(n^{\frac{3}{2}} e^{n\omega_{q,c,d}(\alpha)}\right) \\
\findent \levar{(e)} \order\left(n^{-\ceil{(c-2)l_0/2}}\right) + \order\left(n^{\frac{3}{2}} e^{n\omega_{q,c,d}(\alpha)}\right)
\end{IEEEeqnarray*}
where (a) follows from Markov's inequality, (b) from Theorems~\ref{th:LDPCWeightDistribution} and \ref{th:LDPCWeigthDistributionL0}, Lemma~\ref{le:Inequality1}, and the inequality $l(n-l) \le n^2/4$, (c) from Theorem~\ref{th:PropertyOfOmega}, which shows that $\omega_{q,c,d}(x)$ with $x \in [(l_0+4)/n, \alpha]$ is upper bounded by either $\omega_{q,c,d}((l_0+4)/n)$ or $\omega_{q,c,d}(\alpha)$, (d) from Theorem~\ref{th:InequalityOfOmega}, and (e) follows from $c \ge 3$.

The above inequality holds in all cases. When $q=2$ and $cl_0$ is odd, Theorem~\ref{th:LDPCWeigthDistributionL0} shows that $E[A_{q,c,d}^{(n)}(l_0)] = 0$, so we can further improve this inequality by simply replacing $l_0$ with $l_0+1$. The proof is complete.
\end{IEEEproof}

\begin{remark}
If taking $l_0=1$ in Theorem~\ref{th:MinimumDistance}, we have
\begin{IEEEeqnarray}{l}\label{eq:MinimumDistance2}
P{\left\{\dmin(\mathcal{C}_{q,c,d}^{(n)}) \le n\alpha\right\}} \IEEEnonumber \\
\findent \le \order\left(n^{-\ceil{(c-2)/2}}\right) + \order\left(n^{\frac{3}{2}}e^{n\omega_{q,c,d}(\alpha)}\right).\footnote{When $q=2$ and $c$ is odd, we have a tighter upper bound $\order(n^{2-c}) + \order(n^{\frac{3}{2}}e^{n\omega_{2,c,d}(\alpha)})$. But for simplicity, we ignore this special case.}
\end{IEEEeqnarray}
Recall that $\omega_{q,c,d}(x)$ has a unique zero $x_0(q,c,d)$ in $(0, 1-1/q)$, so we have
\begin{equation}\label{eq:MinimumDistance2a}
P{\left\{\dmin(\mathcal{C}_{q,c,d}^{(n)}) \le n\alpha\right\}} \le \order\left(n^{-\ceil{(c-2)/2}}\right)
\end{equation}
for any $\alpha \in (0, x_0(q,c,d))$. Moreover, when $c \ge 5$, it follows from the Borel-Cantelli lemma that for any $\epsilon>0$, the probability of the event
\[
\left\{\frac{1}{n} \dmin(\mathcal{C}_{q,c,d}^{(n)}) \le x_0(q,c,d) - \epsilon \mbox{ for infinitely many $n$}\right\}
\]
is zero, so that
\begin{equation}
P{\left\{\liminf_{n \to \infty} \frac{1}{n} \dmin(\mathcal{C}_{q,c,d}^{(n)}) \ge x_0(q,c,d)\right\}} = 1.
\end{equation}
The formula \eqref{eq:MinimumDistance2}, for $q = 2$, was first proved (in a slightly stronger form for a different ensemble) by Gallager in \cite{JSCC:Gallager196300}. As for the general case of $q > 2$, Bennatan and Burshtein first showed in \cite{JSCC:Bennatan200403} that there exists some $\gamma > 0$ such that
\[
P{\left\{\dmin(\mathcal{C}_{q,c,d}^{(n)}) \le n\gamma\right\}} \le \order\left(n^{1-c/2}\right)
\]
which is clearly weaker than \eqref{eq:MinimumDistance2a}. In \cite{JSCC:Como2008}, Como and Fagnani proved a result similar to \eqref{eq:MinimumDistance2a}.
\end{remark}

Compared with previous results, the advantage of Theorem~\ref{th:MinimumDistance} is that we can use it to obtain results much better than \eqref{eq:MinimumDistance2} by removing bad codes from the original ensemble. This viewpoint is formulated in the following theorem, which is an easy consequence of Theorem~\ref{th:MinimumDistance}.

\begin{theorem}\label{th:MinimumDistance3}
Let $q\ge 2$, $d \ge c \ge 3$, $l_0 \ge 2$, and $\alpha \in (0, 1-1/q)$. Let $\Phi_n: \{\mbox{All subspaces of $\field{q}^{n}$}\} \to \{0,1\}$ be a test function of linear codes such that for every linear code $C$, $\Phi_n(C) = 1$ implies $\dmin(C) \ge l_0$. If $E[\Phi_{n}(\mathcal{C}_{q,c,d}^{(n)})] \ge \order(\phi(n))$ for some map $\phi(n): \pintegers \to [0,1]$, then
\begin{IEEEeqnarray}{l}
P{\left\{\left.\dmin(\mathcal{C}_{q,c,d}^{(n)}) \le n\alpha \right| \Phi_{n}(\mathcal{C}_{q,c,d}^{(n)}) = 1\right\}} \IEEEnonumber \\
\findent \le \order\left(\frac{n^{-\ceil{(c-2)(l_0+\Delta)/2}}}{\phi(n)}\right) + \order\left(\frac{n^{\frac{3}{2}}e^{n\omega_{q,c,d}(\alpha)}}{\phi(n)}\right) \label{eq:MinimumDistance3}
\end{IEEEeqnarray}
where $\Delta$ is defined by \eqref{eq:DeltaOfL0}.
\end{theorem}

The proof is left to the reader.

\begin{remark}
A simple test function can be defined by checking whether the parity-check matrix of a linear code contains all-zero columns. Then $\Phi_n(C)=1$ if and only if the parity-check matrix of $C$ contains no all-zero columns. It is clear that $\Phi_n(C) = 1$ is equivalent to $\dmin(C) \ge 2$, so it follows from \eqref{eq:MinimumDistance2a} that
\[
E{\left[\Phi_n(\mathcal{C}_{q,c,d}^{(n)})\right]} = P{\left\{\dmin(\mathcal{C}_{q,c,d}^{(n)}) \ge 2\right\}} = \order(1).
\]
Consequently, we have
\begin{IEEEeqnarray}{l}\label{eq:MinimumDistance4}
P{\left\{\left.\dmin(\mathcal{C}_{q,c,d}^{(n)}) \le n\alpha \right| \Phi_{n}(\mathcal{C}_{q,c,d}^{(n)}) = 1\right\}} \IEEEnonumber \\
\findent \le \order\left(n^{2-c}\right) + \order\left(n^{\frac{3}{2}}e^{n\omega_{q,c,d}(\alpha)}\right).
\end{IEEEeqnarray}
\end{remark}

\section{Conclusion}~\label{sec:Conclusion}

We provided a thorough analysis of the average weight distributions of regular LDPC code ensembles over finite fields. The primary results are Theorems~\ref{th:LDPCWeigthDistributionL0}, \ref{th:PropertyOfDelta}, \ref{th:PropertyOfOmega}, and \ref{th:InequalityOfOmega}, which are important for any analysis of regular LDPC codes based on the weight distribution. Furthermore, we proved a general result (Theorem~\ref{th:MinimumDistance}) on the minimum distance of individual codes in a regular LDPC code ensemble, which includes all previous results as special cases.

\section*{Acknowledgment}

The authors would like to thank the anonymous reviewers for their helpful comments.

\appendices

\section{Some Useful Inequalities}\label{appsec:Inequality}

\begin{lemma}\label{le:Inequality1}
For any $n \in \pintegers$, define the function
$$
\beta_{n}(l) \eqdef H_2{\left(\frac{l}{n}\right)} - \frac{1}{n} \ln {n \choose l} \qquad \forall l = 0,1,\ldots, n.
$$
Then
$$
0 \le \beta_{n}(l) \le \frac{1}{2n}\ln\left(\frac{l(n-l)}{n}\right) + \order(n^{-1}) \qquad \forall 0<l<n
$$
and $\beta_n(0)=\beta_n(n)=0$.
\end{lemma}

\begin{IEEEproof}[Sketch of Proof]
Using Stirling's approximation:
\[
n!=\sqrt{2\pi n} \left(\frac{n}{e}\right)^n e^{\lambda_n} \qquad \forall n \ge 1
\]
where $1/(12n+1) < \lambda_n < 1/(12n)$.
\end{IEEEproof}

\begin{lemma}\label{le:Inequality2}
For all $x \in [0, 1]$ and $d \in \pintegers$,
$$
(1 - x)^d \le 1 - dx + \frac{d(d-1)}{2} x^2.
$$
\end{lemma}

\begin{IEEEproof}
The inequality holds trivially for $d = 1$. Now suppose $d \ge 2$, then by Taylor's theorem, it follows that
$$
(1 - x)^d = 1 - dx + \frac{d(d-1)(1-y)^{d-2}}{2} x^2
$$
for some $y \in [0, x]$. This thus concludes the proposition.
\end{IEEEproof}

\section{Derivatives of $H_q(x)$, $D(x\|\hat{x})$, and $\rho_{q,d}(x)$}

\begin{lemma}\label{le:Derivatives}
\begin{IEEEeqnarray*}{rCl}
\frac{dH_q(x)}{dx} &= &\ln \frac{1-x}{x} + \ln(q-1) \\
\frac{\partial D(x\|\hat{x})}{\partial x} &= &\ln \frac{x(1-\hat{x})}{\hat{x}(1-x)} \\
\frac{\partial D(x\|\hat{x})}{\partial \hat{x}} &= &\frac{\hat{x}-x}{\hat{x}(1-\hat{x})} \\
\frac{d\rho_{q,d}(x)}{dx} &= &-\frac{qd\left(\displaystyle 1-\frac{qx}{q-1}\right)^{d-1}}{1 + (q-1) \left(\displaystyle 1-\frac{qx}{q-1}\right)^d}.
\end{IEEEeqnarray*}
where $\rho_{q,d}(x)$ is defined by \eqref{eq:DefinitionOfRho}.
\end{lemma}

The proof is left to the reader.

\section{Descartes' Rule of Signs}

\begin{theorem}[Descartes' Rule of Signs]\label{th:DescartesRule}
If the terms of a univariate polynomial with real coefficients are ordered by ascending or descending variable exponent, then the number of positive roots of the polynomial (counted with their multiplicities) is either equal to the number of sign changes between consecutive nonzero coefficients, or less than it by a multiple of $2$. Since the negative roots of the polynomial equation $f(x) = 0$ are positive roots of the equation $f(-x) = 0$, the rule can be readily applied to help count the negative roots as well.
\end{theorem}

For a proof we refer the reader to \cite{JSCC:Wang200406}.

\section{Proofs of Lemmas in Section~\ref{sec:Function1}}\label{app:Proof1}

\begin{IEEEproof}[Proof of Lemma~\ref{le:PDOfDelta2WithXHat}]
By definition \eqref{eq:DefinitionOfDelta2}, \eqref{eq:PDOfDelta2WithXHat1} follows immediately. Using Lemma~\ref{le:Derivatives} and the change of variables \eqref{eq:TransfromZX} yields
\begin{IEEEeqnarray*}{rCl}
\frac{\partial\delta_{q,d}(x,\hat{x})}{\partial\hat{x}}
&= &\frac{qd(z-\hat{z})}{(1-\hat{z})[1+(q-1)\hat{z}]} - \frac{qd\hat{z}^{d-1}}{1+(q-1)\hat{z}^{d}} \\
&= &\frac{qd\{z-\hat{z}-\hat{z}^{d-1}-[(q-2)-(q-1)z]\hat{z}^d\}}{(1-\hat{z})[1+(q-1)\hat{z}][1+(q-1)\hat{z}^d]} \\
&= &-\frac{qd(\zeta_{q,d}(\hat{z})-z)}{(1-\hat{z})[1+(q-1)\hat{z}]}
\end{IEEEeqnarray*}
as desired.
\end{IEEEproof}

\begin{IEEEproof}[Proof of Lemma~\ref{le:Property1OfZeta}]
To prove the lemma, we have to show that the derivative of $\zeta_{q,d}(\hat{z})$ is continuous on $[-1/(q-1), 1]$ and positive on $(-1/(q-1),1)$. Some tedious manipulation yields
\[
\zeta_{q,d}'(\hat{z}) = \frac{f(\hat{z})}{[1+(q-1)\hat{z}^d]^2}
\]
where
\begin{IEEEeqnarray*}{rCl}
f(\hat{z})
&\eqdef &1+(d-1)\hat{z}^{d-2}+(q-2)d\hat{z}^{d-1}-(q-1)(d-1)\hat{z}^d \\
& &\breakop{-}(q-1)\hat{z}^{2d-2}.
\end{IEEEeqnarray*}
The continuity is obvious, even if $q=2$ and $d$ is odd. Our task is now to show that $f(\hat{z})$ is positive on $(-1/(q-1),1)$. The proof consists of two parts.

First, we show that $f(\hat{z})$ is positive on $[0,1)$. Note that the coefficients of $f(\hat{z})$ have signs $+,+,+,-,-$. By Theorem~\ref{th:DescartesRule} it follows that $f(\hat{z})$ has a unique positive zero. Since $f(0)=1>0$ and $f(1)=0$, it is clear that $f(\hat{z}) > 0$ for all $\hat{z} \in [0,1)$.

Second, we show that $f(\hat{z})$ is also positive on $(-1/(q-1),0)$ for both odd and even $d$.

For odd $d$ we have
\begin{IEEEeqnarray*}{rCl}
f(-\hat{z})
&= &1-(d-1)\hat{z}^{d-2}+(q-2)d\hat{z}^{d-1} \\
& &\breakop{+}(q-1)(d-1)\hat{z}^d-(q-1)\hat{z}^{2d-2}.
\end{IEEEeqnarray*}
If $q\ge 3$ then for all $\hat{z} \in (0,1/(q-1))$,
\begin{IEEEeqnarray*}{rCl}
f(-\hat{z})
&> &1 - (d-1)\hat{z}^{d-2} \\
&> &1 - \frac{(d-1)}{(q-1)^{d-2}} \\
&\ge &\frac{2^{d-1}-(d-1)}{(q-1)^{d-2}} \\
&\ge &0.
\end{IEEEeqnarray*}
As for the case of $q=2$, $f(-\hat{z})$ reduces to
\[
1-(d-1)\hat{z}^{d-2}+(d-1)\hat{z}^d-\hat{z}^{2d-2}
\]
which can be factorized as
\begin{equation}\label{eq:Factorization1}
(1-\hat{z})^3 \left[\sum_{i=0}^{d-3} \frac{(i+1)(i+2)}{2} \left(\hat{z}^{i}+\hat{z}^{2d-5-i}\right)\right]
\end{equation}
so that $f(-\hat{z}) > 0$ for all $\hat{z} \in (0,1)$.

For even $d$ we have
\begin{IEEEeqnarray*}{rCl}
f(-\hat{z})
&= &1+(d-1)\hat{z}^{d-2}-(q-2)d\hat{z}^{d-1} \\
& &\breakop{-}(q-1)(d-1)\hat{z}^d-(q-1)\hat{z}^{2d-2} \\
&> &\hat{z}^{d-2}+(d-1)\hat{z}^{d-2}-\frac{(q-2)d}{q-1}\hat{z}^{d-2} \\
& &\breakop{-}\frac{d-1}{q-1}\hat{z}^{d-2}-\frac{1}{q-1}\hat{z}^{d-2} \\
&= &0
\end{IEEEeqnarray*}
for all $\hat{z} \in (0,1/(q-1))$. The proof is complete.
\end{IEEEproof}

\begin{IEEEproof}[Sketch of Proof of Lemma~\ref{le:Property2OfZeta}]
Identity \eqref{eq:Property2OfXiEq1} is proved by a straightforward argument using definition \eqref{eq:DefinitionOfZeta}. Equations \eqref{eq:Property2OfXiEq2b}, \eqref{eq:Property2OfXiEq3}, and \eqref{eq:Property2OfXiEq4} are immediate consequence of \eqref{eq:Property2OfXiEq1}. As for \eqref{eq:Property2OfXiEq2a}, we note that \eqref{eq:Property2OfXiEq1} with $q=2$ and odd $d$ gives
\[
\zeta_{2,d}(-1)+1 = \left.\frac{z^{d-1}(1-z)}{1-z+z^2-\cdots +z^{d-1}}\right|_{z=-1} = \frac{2}{d}
\]
so that $\zeta_{2,d}(-1) = 2/d-1$.
\end{IEEEproof}

\begin{IEEEproof}[Proof of Lemma~\ref{le:Property3OfZeta}]
Lemmas~\ref{le:Property1OfZeta} and \ref{le:Property2OfZeta} show that the range of $\zeta_{q,d}(\hat{z})$ for $\hat{z} \in [-1/(q-1),1]$ is
\[
\left[\zeta_{q,d}{\left(-\frac{1}{q-1}\right)}, \zeta_{q,d}(1)\right] = [z_1, 1]
\]
and therefore the equation $\zeta_{2,d}(\hat{z})-z=0$ has a unique solution in $[-1/(q-1),1]$ for each $z\in [z_1,1]$  and has no solution in $[-1/(q-1),1]$ for $z < z_1$.

Since $\zeta_{q,d}(\hat{z})$ is continuously differentiable on $[-1/(q-1), 1]$ and its derivative is positive on $(-1/(q-1),1)$, it follows from the inverse function theorem that the solution $\hat{z}_1(z)$ is continuously differentiable on $(z_1, 1)$ and its derivative is also positive on $(z_1, 1)$. The continuity of $\hat{z}_1(z)$ at endpoints also follows. Moreover, Lemma~\ref{le:Property2OfZeta} shows that
\begin{IEEEeqnarray*}{ll}
\zeta_{q,d}(z)=z_1 &\qquad\mbox{if $z = -\frac{1}{q-1}$} \\
\zeta_{q,d}(z)>z &\qquad\mbox{if $z \in (-\frac{1}{q-1},0)$ and $d$ is odd} \\
\zeta_{q,d}(z)<z &\qquad\mbox{if $z \in (-\frac{1}{q-1},0)$ and $d$ is even} \\
\zeta_{q,d}(z)=0 &\qquad\mbox{if $z=0$} \\
\zeta_{q,d}(z)>z &\qquad\mbox{if $z \in (0,1)$} \\
\zeta_{q,d}(z)=1 &\qquad\mbox{if $z=1$}.
\end{IEEEeqnarray*}
This implies that $\hat{z}_1(z) \in I_{q,d}'(z)$.
\end{IEEEproof}

\section{Proofs of Lemmas in Section~\ref{sec:Function2}}\label{app:Proof2}

\begin{IEEEproof}[Proof of Lemma~\ref{le:D1OfOmega}]
Definition \eqref{eq:DefinitionOfOmega} and equation \eqref{eq:PropertyEq2OfDelta} show that
\begin{IEEEeqnarray*}{rCl}
\frac{d\omega_{q,c,d}(x)}{dx}
&= &\frac{dH_q(x)}{dx} + c \ln \frac{x(1-\hat{x}_1)}{\hat{x}_1(1-x)} \\
&\eqvar{(a)} &\ln \frac{1-x}{x} + \ln(q-1) + c \ln \frac{x(1-\hat{x}_1)}{\hat{x}_1(1-x)} \\
&= &\ln \left[\left(\frac{x}{1-x}\right)^{c-1} \left(\frac{1-\hat{x}_1}{\hat{x}_1}\right)^{c}\right] + \ln(q-1)
\end{IEEEeqnarray*}
where (a) follows from Lemma~\ref{le:Derivatives}. By Lemma~\ref{le:PDOfDelta2WithXHat}, equation \eqref{eq:DefinitionOfXHat1} is equivalent to $\zeta_{q,d}(\hat{z}_1) - z = 0$, where $z = 1 - qx/(q-1)$ and $\hat{z}_1 = 1 - q\hat{x}_1/(q-1)$. After some manipulations, we obtain
\[
\frac{x}{\hat{x}_1}
= \frac{1-z}{1-\hat{z}_1}
= \frac{1-\hat{z}_1^{d-1}}{1+(q-1)\hat{z}_1^d}
\]
and
\[
\frac{1-x}{1-\hat{x}_1}
= \frac{1+(q-1)z}{1+(q-1)\hat{z}_1}
= \frac{1+(q-1)\hat{z}_1^{d-1}}{1+(q-1)\hat{z}_1^d}.
\]
Then
\begin{IEEEeqnarray*}{rCl}
\frac{d\omega_{q,c,d}(x)}{dx}
&= &\ln \left\{\frac{(q-1)(1-\hat{x}_1)}{\hat{x}_1} \left[\frac{x(1-\hat{x}_1)}{\hat{x}_1(1-x)}\right]^{c-1}\right\} \\
&= &\ln \left\{\frac{1+(q-1)\hat{z}_1}{1-\hat{z}_1} \left[\frac{1-\hat{z}_1^{d-1}}{1+(q-1)\hat{z}_1^{d-1}}\right]^{c-1}\right\}.
\end{IEEEeqnarray*}
The proof is complete.
\end{IEEEproof}

\begin{IEEEproof}[Proof of Lemma~\ref{le:D1OfOmegaAtSpecialPoints}]
From Theorem~\ref{th:PropertyOfDelta}, it follows  that $\lim_{x\to 0^+} \hat{z}_1 = 1$. Then equation \eqref{eq:D1OfOmega2} with $c=1$ and $c \ge 3$ gives \eqref{eq:D1OfOmegaAtSpecialPoints1a} and \eqref{eq:D1OfOmegaAtSpecialPoints1c}, respectively. As for $c=2$, we have
\begin{IEEEeqnarray*}{l}
\lim_{x\to 0^+} \frac{d\omega_{q,c,d}(x)}{dx} \\
\findent = \lim_{\hat{z}_1 \to 1^-} \ln \left\{\frac{[1+(q-1)\hat{z}_1](1-\hat{z}_1^{d-1})}{(1-\hat{z}_1)[1+(q-1)\hat{z}_1^{d-1}]}\right\} \\
\findent = \lim_{\hat{z}_1 \to 1^-} \ln \left\{\frac{[1+(q-1)\hat{z}_1](1+\hat{z}_1+\cdots+\hat{z}_1^{d-2})}{1+(q-1)\hat{z}_1^{d-1}}\right\} \\
\findent = \ln(d-1).
\end{IEEEeqnarray*}
By the symmetric property (Remark~\ref{re:SymmetricProperty}), we also obtain \eqref{eq:D1OfOmegaAtSpecialPoints3}.

From Theorem~\ref{th:PropertyOfDelta}, it follows  that
\[
\hat{z}_1{\left(1-\frac{1}{q}\right)} = 1 - \frac{q(1-1/q)}{q-1} = 0.
\]
This together with equation \eqref{eq:D1OfOmega2} gives \eqref{eq:D1OfOmegaAtSpecialPoints2}.

Again by Theorem~\ref{th:PropertyOfDelta}, it follows  that $\lim_{x\to x_1^-} \hat{z}_1 = -1/(q-1)$. Then \eqref{eq:D1OfOmega2} with $q \ne 2$ or $d$ odd gives \eqref{eq:D1OfOmegaAtSpecialPoints4}.
\end{IEEEproof}

\begin{IEEEproof}[Proof of Lemma~\ref{le:D2ZOfOmega}]
It follows from Lemma~\ref{le:D1OfOmega} that
\begin{IEEEeqnarray*}{l}
\frac{d}{d\hat{z}_1} {\left(\frac{d\omega_{q,c,d}(x)}{dx}\right)} \\
\findent = \frac{q}{[1+(q-1)\hat{z}_1](1-\hat{z}_1)} - \frac{q(c-1)(d-1)\hat{z}_1^{d-2}}{(1-\hat{z}_1^{d-1})[1+(q-1)\hat{z}_1^{d-1}]} \\
\findent = \frac{q\xi_{q,c,d}(\hat{z})}{(1-\hat{z}_1^{d-1})[1+(q-1)\hat{z}_1][1+(q-1)\hat{z}_1^{d-1}]}.
\end{IEEEeqnarray*}
This concludes \eqref{eq:D2ZOfOmega1}, while the first equality with $c=1$ gives \eqref{eq:D2ZOfOmega2}.
\end{IEEEproof}

\begin{IEEEproof}[Proof of Lemma~\ref{le:PropertyOfXi}]
Since $\xi_{q,c,d}(0)=1$, it suffices to determine all zeros of $\xi_{q,c,d}(\hat{z})$ in $(-1/(q-1), 1)$. The proof consists of two parts.

First, we check the zeros of $\xi_{q,c,d}(\hat{z})$ in $(0, 1)$. We note that the coefficients of $\xi_{q,c,d}(\hat{z})$ have signs $+,\ldots, +,-,\linebreak[0] -,+, \ldots, +$. By Theorem~\ref{th:DescartesRule} it follows that $\xi_{q,c,d}(\hat{z})$ has zero or two positive zeros. On the other hand,
\[
\xi_{q,c,d}(0)=1,\; \xi_{q,c,d}(1)=-q(c-2)(d-1), \; \xi_{q,c,d}(\infty)=\infty
\]
and
\[
\xi_{q,2,d}'(1) = \frac{1}{2}(q-2)(d-1)(d-2).
\]
Then for $q \ge 2$ and $d \ge c \ge 3$, $\xi_{q,c,d}(\hat{z})$ has a unique zero $\hat{z}_2$ in $(0, 1)$. As for $c=2$, $\xi_{q,2,d}(\hat{z})$ with $q\ge 3$ has a unique zero $\hat{z}_2$ in $(0,1)$ since $\xi_{q,2,d}(1)=0$ and $\xi_{q,2,d}'(1)>0$, while $\xi_{2,2,d}(\hat{z})$ has only one zero $\hat{z}=1$ in $(0, \infty)$ since $\xi_{2,2,d}'(1)=0$ (a zero of multiplicity $2$), so that $\xi_{2,2,d}(\hat{z})$ is positive on $(0,1)$.

Second, we check the zeros of $\xi_{q,c,d}(\hat{z})$ in $(-1/(q-1), 1)$. To facilitate the analysis, we consider the function
\[
f_{q,c,d}(\hat{z}) \eqdef (1+\hat{z})\xi_{q,c,d}(-\hat{z}).
\]
Then the zeros of $\xi_{q,c,d}(\hat{z})$ in $(-1/(q-1), 0)$ are just the zeros of $f_{q,c,d}(\hat{z})$ in $(0, 1/(q-1))$.

If $d$ is odd, we have
\begin{IEEEeqnarray*}{rCl}
f_{q,c,d}(\hat{z})
&= &(1-\hat{z}^{d-1})[1+(q-1)\hat{z}^{d-1}] \\
& &\breakop{+} (c-1)(d-1)\hat{z}^{d-2} (1+\hat{z})[1-(q-1)\hat{z}]
\end{IEEEeqnarray*}
which is clearly positive for all $\hat{z} \in (0, 1/(q-1))$.

If $d$ is even, we have
\begin{IEEEeqnarray*}{rCl}
f_{q,c,d}(\hat{z})
&= &(1+\hat{z}^{d-1})[1-(q-1)\hat{z}^{d-1}] \\
& &\breakop{-} (c-1)(d-1)\hat{z}^{d-2} (1+\hat{z})[1-(q-1)\hat{z}] \\
&= &1 - (c-1)(d-1)\hat{z}^{d-2} \\
& &\breakop{+} (q-2)[(c-1)(d-1)-1] \hat{z}^{d-1} \\
& &\breakop{+} (q-1)(c-1)(d-1)\hat{z}^d - (q-1)\hat{z}^{2d-2}.
\end{IEEEeqnarray*}
When $q=2$, it reduces to
\[
f_{2,c,d}(\hat{z}) = 1 - (c-1)(d-1)\hat{z}^{d-2} + (c-1)(d-1)\hat{z}^d - \hat{z}^{2d-2}.
\]
Since the coefficients of $f_{2,c,d}(\hat{z})$ have signs $+,-,+,-$, it follows from Theorem~\ref{th:DescartesRule} that $f_{2,c,d}(\hat{z})$ has one or three positive zeros. Moreover, we note that
\[
f_{2,c,d}(0) = 1,\; f_{2,c,d}(1) = 0,\; f_{2,c,d}(\infty) = -\infty
\]
and
\[
f_{2,c,d}'(1) = 2(c-2)(d-1).
\]
Then $f_{2,c,d}(\hat{z})$ with $c \ge 3$ has a unique zero $\hat{z}_2'$ in $(0,1)$, while $f_{2,2,d}(\hat{z})$ is positive on $(0,1)$ because of \eqref{eq:Factorization1}. Finally, let us show that $f_{q,c,d}(\hat{z})$ is positive on $(0, 1/(q-1))$ for $q\ge 3$, $c \ge 2$, and $d \ge \max\{c, 4\}$. Since $q \ge 3$, $c \ge 2$, $d \ge 4$, and $\hat{z} < 1/(q-1)$,
\begin{IEEEeqnarray}{rCl}
f_{q,c,d}(\hat{z})
&> &1 - (c-1)(d-1)\hat{z}^{d-2}(1-\hat{z}-2\hat{z}^2) \IEEEnonumber \\
& &\breakop{-} \hat{z}^{d-1} - \hat{z}^{2d-3} \label{eq:Eq1InProofOfXiLem} \\
&> &1 - (c-1)(d-1)\hat{z}^{d-2}. \label{eq:Eq2InProofOfXiLem}
\end{IEEEeqnarray}
For $\hat{z} \in (0, \frac{1}{3}]$, inequality \eqref{eq:Eq2InProofOfXiLem} shows that
\[
f_{q,c,d}(\hat{z}) > 1 - (d-1)^2\left(\frac{1}{3}\right)^{d-2} \ge 1 - (4-1)^2 \left(\frac{1}{3}\right)^{4-2} = 0.
\]
For $\hat{z} \in (\frac{1}{3}, \frac{2}{5}]$, inequality \eqref{eq:Eq1InProofOfXiLem} shows that
\begin{IEEEeqnarray*}{rCl}
f_{q,c,d}(\hat{z}) &> &1 - \frac{4(d-1)^2}{9}\left(\frac{2}{5}\right)^{d-2} - \left(\frac{2}{5}\right)^{d-1} - \left(\frac{2}{5}\right)^{2d-3} \\
&\ge &1 - \frac{4}{9} \cdot 3^2\left(\frac{2}{5}\right)^{2} - \left(\frac{2}{5}\right)^{3} - \left(\frac{2}{5}\right)^{5} \\
&= &\frac{893}{3125}.
\end{IEEEeqnarray*}
For $\hat{z} \in (\frac{2}{5}, \frac{1}{2})$, inequality \eqref{eq:Eq1InProofOfXiLem} shows that
\begin{IEEEeqnarray*}{rCl}
f_{q,c,d}(\hat{z}) &> &1 - \frac{7(d-1)^2}{25}\left(\frac{1}{2}\right)^{d-2} - \left(\frac{1}{2}\right)^{d-1} - \left(\frac{1}{2}\right)^{2d-3} \\
&\ge &1 - \frac{7}{25} \cdot 3^2\left(\frac{1}{2}\right)^{2} - \left(\frac{1}{2}\right)^{3} - \left(\frac{1}{2}\right)^{5} \\
&= &\frac{171}{800}.
\end{IEEEeqnarray*}
The proof is complete.
\end{IEEEproof}


\begin{biographynophoto}{Shengtian Yang}
(S'05--M'06) was born in Hangzhou, Zhejiang, China, in 1976. He received the B.S. and M.S. degrees in biomedical engineering, and the Ph.D. degree in electrical engineering from Zhejiang University, Hangzhou, China in 1999, 2002, and 2005, respectively.

From June 2005 to December 2007, he was a Postdoctoral Fellow at the Department of Information Science and Electronic Engineering, Zhejiang University. From December 2007 to January 2010, he was an Associate Professor at the Department of Information Science and Electronic Engineering, Zhejiang University. Currently, he is a self-employed Independent Researcher in Hangzhou, China. His research interests include information theory, coding theory, and design and analysis of algorithms.
\end{biographynophoto}

\begin{biographynophoto}{Thomas Honold}
(M'95) was born in Munich, Germany, in 1962. He received
his Diplom (1990), doctoral degree (1994) and
Habilitation (2000, the qualification for university teaching in
Germany) in Mathematics from TU Munich, Germany.  He has held
appointments at TU Munich, University of Eichst\"att, Germany, and the
German Institute of Science and Technology, Singapore. Since 2007 he
is working as Associate Professor for the Department of Information Science
and Electronic Engineering, Zhejiang University, Hangzhou. His main
research interest is coding theory and geometry over finite fields and
rings.
\end{biographynophoto}

\begin{biographynophoto}{Yan Chen}
(S'06--M'10) was born in Hangzhou, Zhejiang, China, in 1982. She received the B.Sc. and the Ph. D degree in information and communication engineering from Zhejiang University, Hangzhou, China, in 2004 and 2009, respectively. She has been a Visiting Researcher at the Department of Electronic and Computer Engineering, Hong Kong University of Science and Technology, Hong Kong. After graduation, she joined Huawei Technologies (Shanghai) Co., Ltd. and is currently working as a Research Engineer in the Central Research Department. Her current research interests include green network information theory, energy-efficient network architecture and management, fundamental tradeoffs on green wireless network design, as well as the radio technologies and resource allocation optimization algorithms therein.
\end{biographynophoto}

\begin{biographynophoto}{Zhaoyang Zhang}
(M'02) was born in Huanggang, Hubei, China, in 1973. He received the B.Sc. degree in radio technology and the Ph.D degree in information and communication engineering from Zhejiang University, Hangzhou, China, in 1994 and 1998, respectively.

Since 1998, he has been with the Department of Information Science and Electronic Engineering, Zhejiang University, Hangzhou, China, where he is currently a Professor. His current research interests include information theory and signal processing theory with emphsis on their applications in wireless communications and networks.
\end{biographynophoto}

\enlargethispage{-8.2in}

\begin{biographynophoto}{Peiliang Qiu}
(M'03) was born in Shanghai, China, in 1944. He received the B.S. degree from the Harbin Institute of Technology, Harbin, China, in 1967 and the M.S. degree from the Graduate School of Chinese Academy of Science, Beijing, in 1981, both in electronics engineering.

From 1968 to 1978, he was a Research Engineer at Jiangnan Electronic Technology
Institute. Since November 1981, he has been with Zhejiang University, Hangzhou, China, where he is currently a Professor at the Department of Information Science and Electronic Engineering. His current research interests include digital communications, information theory, and wireless networks.
\end{biographynophoto}

\end{document}